\documentclass[useAMS,usenatbib]{mn2e}

\usepackage{amssymb} 
\usepackage{epsfig} 
\usepackage{natbib}

\def\I{{\em INTEGRAL}}
\def\R{{\em RXTE}} 
\def\S{{\em Swift}}
\def\IGR{IGR~J17473-2721} 
\def\gcm2{g~cm$^{-2}$}
\def\gcms{g~cm$^{-2}$~s$^{-1}$} 
\def\ergcs{erg~cm$^{-2}$~s$^{-1}$} 
\def\ergcm{erg~cm$^{-2}$} 
\def\ergps{erg~s$^{-1}$}

\begin{document} 
 
\title[IGR~J17473-2721 burst behaviour]
{Puzzling thermonuclear burst behaviour from the transient low-mass X-ray binary IGR~J17473-2721}
 
\author[J.Chenevez et al]
{J.~Chenevez,$^1$\thanks{E-mail: jerome@space.dtu.dk} 
D.~Altamirano,$^2$ 
D.K.~Galloway,$^3$\thanks{also School of Physics, Monash University, Australia}
J.J.M.~in~'t~Zand,$^4$ 
E.~Kuulkers,$^5$  
\newauthor 
N.~Degenaar,$^2$ 
M.~Falanga,$^6$ 
E.~Del Monte,$^7$  
Y.~Evangelista,$^7$ 
M.~Feroci,$^7$ 
E.~Costa$^7$ \\
$^1$National Space Institute, Technical University of Denmark, 
Juliane Maries Vej 30, 2100 Copenhagen Ø, Denmark \\
$^2$Astronomical Institute, "Anton Pannekoek", University of Amsterdam, 
Science Park 904, 1098 XH Amsterdam, The Netherlands \\
$^3$Center for Stellar and Planetary Astrophysics, Monash University, VIC 3800, Australia \\
$^4$SRON Netherlands Institute for Space Research, 
Sorbonnelaan 2, 3584 CA Utrecht, The Netherlands  \\
$^5$ISOC, ESA/ESAC, Urb. Villafranca del Castillo, P.O. Box 50727, E-28080 Madrid, Spain \\ 
$^6$International Space Science Institute, Hallerstrasse 6, CH-3012 Bern, Switzerland \\
$^7$INAF/IASF, via Fosso del Cavaliere 100, I-00133, Roma, Italy \\
} 
 
\maketitle

\begin{abstract}
We investigate the thermonuclear bursting behaviour of \IGR,
an X-ray transient that in 2008 underwent a six month long outburst,  
starting (unusually) with an X-ray burst.
We detected a total of 57 thermonuclear bursts throughout the outburst with 
{\it AGILE}, \S, \R, and \I.
The wide range of inferred accretion rates (between $<1\%$ and $\simeq 20\%$ of 
the Eddington accretion rate $\dot m_{\rm Edd}$)
spanned during the outburst allows us to study changes in the nuclear burning 
processes and to identify up to seven different phases. 
The burst rate increased gradually with the accretion rate until it 
dropped (at a persistent flux corresponding to $\simeq 15\%$ of $\dot m_{\rm Edd}$) 
a few days before the outburst peak, after which bursts were not detected for a month. 
As the persistent emission subsequently decreased, the bursting activity 
resumed at a much lower rate than during the outburst rise.
This hysteresis may arise from the thermal effect of 
the accretion on the surface nuclear burning processes, and the timescale
is roughly consistent with that expected for the neutron star crust thermal response.
On the other hand, an undetected ``superburst'', occurring within a data gap near 
the outburst peak, could have produced a similar quenching of burst\,activity.
\end{abstract}


\section{Introduction} 
\label{sec:intro} 

X-ray bursters are accreting neutron stars (NS) in low-mass X-ray binary (LMXB) systems, 
in which hydrogen (H) and helium (He) accumulates on the surface, 
periodically exploding in thermonuclear runaways. These thermonuclear flashes, 
observed as 
type~I X-ray bursts \citep[e.g.,][and hereafter simply called X-ray bursts]{Lew93}, 
are caused by the high temperatures and densities 
reached at the base of the fuel layer.
Such X-ray bursts are characterised by black-body emission with peak 
temperature $kT\simeq 2-3$~keV and a light curve showing a fast rise and exponential decay 
\citep[for reviews see][]{Lew93, StrohBil06}.
To date, a total of 92 X-ray bursters have been identified in the Milky Way 
\citep[e.g.,][]{intZ04, Liu07}\,\footnote{http://www.sron.nl/$\sim$jeanz/bursterlist.html}. 
Many burst sources exhibit ``photospheric radius expansion'' (PRE) bursts, 
attributed to radiation pressure exceeding the gravitational
attraction in the photosphere. 
Strong variations of the black-body radius are then observed simultaneously with 
inverse variations of the colour temperature at a constant luminosity,
roughly equal to the Eddington limit ($L_{\rm Edd}$). 
Such bursts may thus be used to estimate the distance to the burst sources 
\citep[e.g.][]{Basinska84}.

The unstable thermonuclear burning is a recurrent phenomenon for which ignition 
conditions depend primarily on the (local) mass accretion rate.
Early theoretical studies \citep{Fuji81, Fushi87} predict three distinct nuclear 
burning regimes as a function of increasing accretion rate per unit area $\dot m$ 
of H/He-rich material with solar metallicity \citep[see also][]{StrohBil06}: 
unstable H ignition triggers He flashes for $\dot m \lesssim 900~$ \gcms; 
pure He flashes occur below a shell of H steadily burning in the hot CNO cycle 
for $900 \lesssim \dot m \lesssim 2000~$ \gcms; 
and mixed He/H bursts are triggered by unstable He ignition for $\dot m \gtrsim 2000~$ \gcms.
Still, the detailed physics of the nuclear burning are not completely understood.
The above-mentioned thresholds depend on nuclear reactions rates, sedimentation 
and mixing processes in the burning layers, as well as anisotropy effects. 
Thus, some phenomenological issues still have to be solved, 
such as the drop in burst activity at high accretion rate (e.g. Cornelisse et al. 2003), 
as well as the precise burst mechanism at low accretion rates $\lesssim 1\%~\dot{m}_{\rm Edd}$.

Due to nuclear reaction waiting points related to a series of $\beta$ decays 
during the burning of H in the rp-process, H-dominated bursts burn slower 
than He-dominated bursts, that burn instead through the fast triple-$\alpha$ reactions. 
The lengths of the rise times and exponential decays are therefore related to the 
relative amounts of H to He burnt during the bursts.
However, bursts lasting several tens of minutes have also been recorded, 
that are now interpreted as the long burning of a thick pure He layer,
accumulated over a long interval at very low accretion rate and/or from an He donor 
\citep[see, e.g.,][]{cum06, intZ05, chenevez07, MF08}.
Even longer and more energetic nuclear bursts have been reported in rare occasions 
\citep[see, e.g.,][]{KeekZand08} as superbursts that last up to several hours. 
They are thought to arise from carbon (C) burning in a thick layer below the surface 
of the NS \citep{CB01, StrohBrow02}. 
Although H/He burning is likely required to produce the C for these
bursts, interestingly the occurrence of a superburst has the effect of
'quenching' normal bursting activity for weeks to months \citep{Kuul04}.

X-ray transients are ideal sources for studies of burst behaviour
because they frequently experience a large range of accretion rates (and spectral states)
in a short time frame.
The X-ray transient source \IGR\ 
was discovered with 
the INTErnational Gamma-Ray Astrophysics Laboratory (\I) in the Galactic Centre 
region during an outburst in April 2005 \citep{Greb05}. 
The source XTE J1747-274, detected by \R\ in May 2005 \citep{MarkSwank05}, was rapidly 
identified with \IGR\ by \citet{Kennea05} using \S/XRT.
The source exhibited a second episode of activity between March and September 2008, 
which 
began with a type~I X-ray burst on MJD~54552 
\citep{DelMonte}, thus identifying the source as a neutron star in a LMXB system 
\citep[though the first bursts ever observed from this source are reported in][hereafter G08]{Galloway08}. 
Only a few observations of thermonuclear flashes occurring before the start of an outburst 
have been recorded \citep[e.g.,][]{cor07, Kuul09}.  

The 2008 outburst of \IGR\ was well followed from its beginning by most 
of the high energy satellite fleet operating at that time. 
{\it Swift}/XRT \citep{Alta1459} and \R/PCA \citep{Mark1460} confirmed renewed 
accretion activity from \IGR\ at a position consistent with the burst detection by 
SuperAGILE \citep{DelMonte}. 
Hard X-ray persistent emission was almost simultaneously detected by \I\ 
\citep{kuul1461, Bal1468}.
Later on during the outburst episode, kilohertz QPOs at frequencies up to 
$\simeq 900$~Hz were detected by \citet{Alta1651}, 
after a spectral switch from a hard to a soft state. 
The same authors thus identified the source to belong to the class of atoll NS, 
and, from the observation of two PRE bursts, estimated its distance 
to be in the range
4.9--5.7~kpc. 
A detailed comparison of the \IGR\ outbursts in 2005 and 2008 is described by \citet{Zhang09}, 
and a study of some bursts observed in 2008 is presented by the same authors in \citet{Chen2010} 
that we discuss in section \ref{subsec:Summary}.

In the present paper, we investigate 
the bursting behaviour of \IGR\ during its outburst in 2008, so all calendar dates refer to that year. 
This study is based on all the bursts observed by instruments aboard {\em AGILE}, \S, \R, and \I.

\section{Observations and data analysis}
\label{sec:data}

\subsection{AGILE}
\label{subsec:SuperAGILE}  
The {\it AGILE}\/ satellite mission \citep{Tavani08}, launched on 23 April 2007, 
includes a hard X-ray (18--60 keV) monitor, SuperAGILE \citep[see][2010, for details]{Feroci07}, 
that consists of two pairs of 1-D coded aperture imagers of 
$107^{o} \times 68^{o}$ field of view (full width zero response), and pointed to the same target. 
Two are oriented along one direction in the sky and the other two in the orthogonal direction. 
In the overlapping region, of $68^{o} \times 68^{o}$, both coordinates are encoded, thus 
giving twice a 1-D imaging, and the sensitivity on axis is of the order of 15 mCrab at $5 \sigma$ 
in one day.

SuperAGILE is equipped with ground trigger software \citep[see][for details]{DelMonte08}, 
dedicated to the detection of $\gamma$-ray and X-ray bursts, as well as other transients, 
occurring on time scales from 512 to 8192~ms. The trigger sensitivity 
in 10~s at 5 sigma corresponds to about 1.3 Crab assuming a Crab-like spectrum, 
or to $4.8 \times 10^{-9}$~\ergcs, assuming a 3 keV black-body spectrum. 
In case of a trigger, an image is extracted and the burst position can be reconstructed 
with an uncertainty of 3 arcmin radius.

\subsection{Swift}
\label{subsec:Swift}
The NASA mission \S\ \citep{Gehrels04} provides 15--50 keV light curves using the 
Burst Alert Telescope\footnote{http://swift.gsfc.nasa.gov/docsd/swift/results/transients} 
\citep[BAT;][]{Barthelmy05}, 
with which we followed the long-term hard X-ray evolution of the outburst of \IGR.

Following the 
burst detection by SuperAGILE (see section \ref{subsec:bursts}), \cite{Alta1459} 
obtained a \S\ observation to conclusively identify the burst source with \IGR\ 
on March 31 (MJD~54556). 
We analysed data from the X-Ray Telescope \citep[XRT; 0.2--10 keV,][]{Burrows05} 
in both Photon Counting (PC) and Windowed Timing (WT) modes, 
resulting in exposure times of 4.1 and 0.7 ksec respectively.  
The raw data were processed with the XRT pipeline using standard quality cuts 
(event grades 0-2 and 0-12 for the WT and PC mode respectively). 
The PC spectrum was fit with \texttt{XSpec} version~12 \citep{arnaud96} 
to a simple absorbed power-law model of index 1.85 with a reduced $\chi^2=1.24$ for 187 dof. 

During the portion of the observation for which WT data was available, 
a type~I X-ray burst was detected with a peak count rate above 400~counts~s$^{-1}$.  
We carried out time-resolved spectroscopy by extracting source events
from a rectangular box of 40 pixels 
along the image strip and 20 pixels wide.
According to \citet{Romano06}, WT data at such high count rates are expected to be affected by 
pile-up, so we excluded a box of the central 4 by 20 pixels for intervals in which the count rate 
exceeded 400~counts~s$^{-1}$, 2 by 20 pixels for rates of 300--400~counts~s$^{-1}$, 
and a band of 1 by 20 pixels for count rates between 100--300~counts~s$^{-1}$. 
The burst time resolved spectral analysis was performed dividing up the burst in seven intervals from 
2~s to 100~s. 
These data were fit using \texttt{XSpec} with a black-body model subject to interstellar absorption, 
fixing the hydrogen column density to the value found from fitting the PC spectral data 
($N_{\rm H}=4.39\pm 0.35 \times 10^{22}$~cm$^{-2}$).

\subsection{INTEGRAL}
\label{subsec:INTEGRAL}
The hard X-ray and $\gamma$-ray observatory \I\ \citep{w03} provides  
three 
coded-mask instruments operating simultaneously.  
For our study we used the X-ray monitor JEM-X \citep[3--25 keV,][]{lund03} 
and IBIS/ISGRI \citep[18~keV--10~MeV,][]{lebr03} 
data, reduced with the standard Off-line Science Analysis \citep[OSA;][]{c03} 
software version 8.0. 
Data were corrected for vignetting effects of the JEM-X collimator. 
A systematic error of 3\% per channel was applied to JEM-X and ISGRI spectra, 
corresponding to the current estimated uncertainties in their response.
Due to the partly incomplete coded-mask pattern of JEM-X, its signal deconvolution 
can be affected by some crosstalk between sources inside the same field of view. 
We have verified that the source-extracted signal for the analysis of \IGR\ burst events 
was not contaminated 
by neighbouring sources\footnote{In
particular GX~3+1, located 0.8$^{\circ}$ away from \IGR\, 
and which at that time was in a high-soft, not bursting state.}.

The outburst of \IGR\ was monitored as part of \I\ regular observations 
of the Galactic Bulge performed each spring and fall, approximately every 
three days for about 3.5~h \citep{GB07}. 
In 2008, this monitoring programme covered \IGR\ between February 11 (MJD~54507) 
and April 20 (MJD~54516), 
and again between August 18 (MJD~54696) and October 22 (MJD~54761). 
More \I\ data originate from long observations of the Galactic Centre region performed on April 6 (MJD~54562). 
We searched for X-ray bursts by scrutinizing the light curves obtained from every pointing 
within $\leq 5^{\circ}$ (JEM-X) and $\leq 12^{\circ}$ (ISGRI) of the source position,
for a total effective exposure 
of about 500~ks and 940~ks, respectively.

\begin{figure*} 
\centerline{\epsfig{file=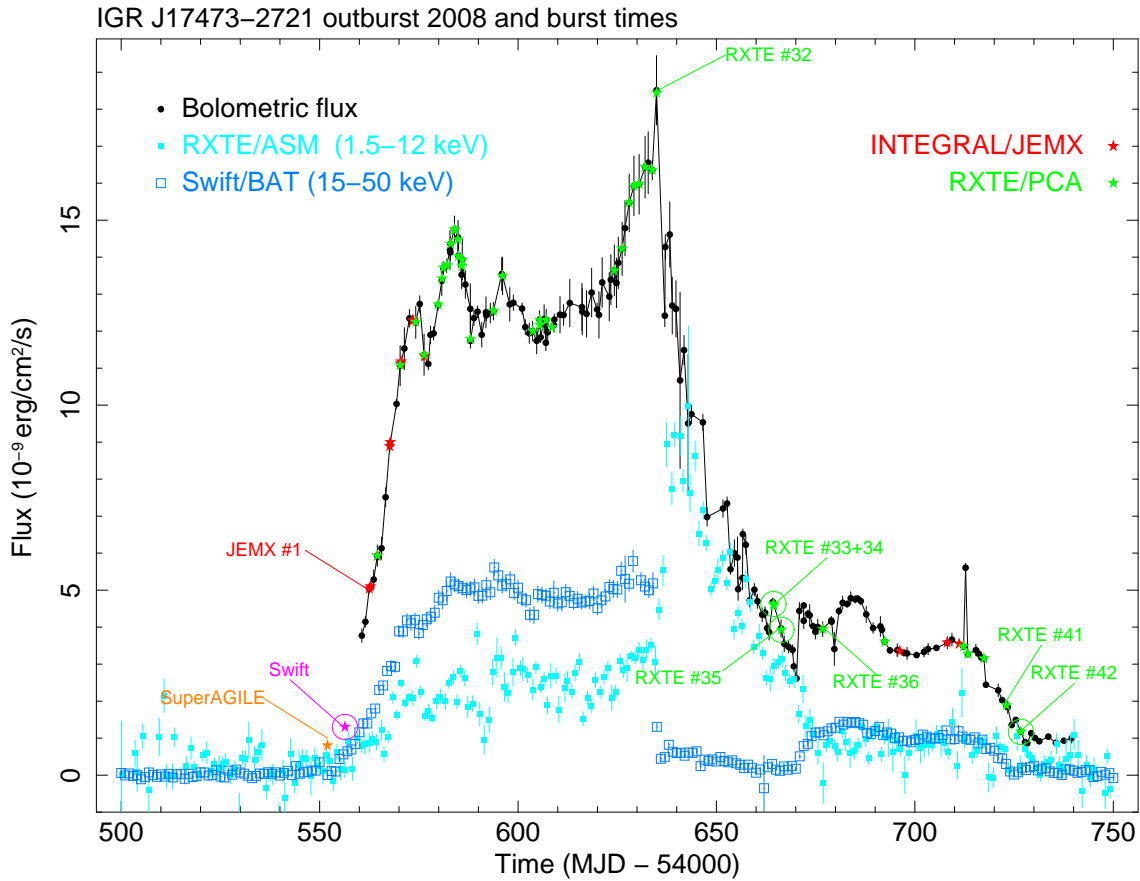,angle=-90,width=15cm}} 
\caption 
{
Broad band X-ray light curves for \IGR\ during the 2008 outburst,
combining all available \R\ and \S/BAT 1-day averaged data.
The bolometric fluxes are derived from \R/(PCA+HEXTE) fits (see text) and have been connected for clarity. 
Every PCA observation is represented by a black dot. 
The observed X-ray bursts are represented by stars, with circles for the four PRE ones, 
and those indicated by names delineate the transitions between different
phases of burst behaviour 
(see sections \ref{subsec:burstacc} and \ref{subsec:Summary}).
}  
\label{fig:broad_outburst} 
\end{figure*}

\subsection{RXTE}
\label{subsec:RXTE}

We used data taken by the three instruments on board NASA's 
{\it Rossi X-ray Timing Explorer} ({\it RXTE}). 
The densest coverage of the outburst at 2.5--12 keV is provided by the 
All Sky Monitor \citep[ASM;][]{l96} publicly available light curves.

The Proportional Counter Array \citep[PCA;][]{Jahoda} consists of five large area 
Proportional Counter Units (PCUs). 
We used data from PCU2 as it was always on during the relevant observations. 
The PCA data set is composed of two independent series of observations between 
April 4 (MJD~54560) and May 22 (MJD~54608), 
as well as regular public observations 
from May 23 (MJD~54609) to October 2 (MJD~54741).
The pointing was offset by $0.25^{\circ}$ in order to avoid GX~3+1 falling inside the 
$\approx1^{\circ}$ PCA field of view,
and no other nearby burst sources 
were active during that period.
The {\it RXTE}/PCA observations were performed about every day during the 
outburst,
with typically $\simeq 2.6$~ks non-interrupted exposure each time.
The total source exposure is $490$~ks.

We followed the evolution of the persistent emission during the outburst by comparing, 
for each PCA observation, its intensity, defined as the count rate in the 2--16 keV energy band, 
with its hardness, defined as the ratio of the intensity in the 9.7--16 keV band 
to that in the 6--9.7 keV band. 
We thus drew a hardness-intensity diagram (HID) where these quantities are normalised 
to the corresponding values obtained with the Crab 
\citep[see][for a full description of the procedure]{Alta2008}.
We fitted the vignetting-corrected PCU2 spectra of the 2.5--25~keV persistent emission of \IGR\ 
with a model consisting of a black-body radiation plus a power-law and a gaussian emission line 
at 6.4~keV (consistent with FeI K$\alpha$ fluorescence) all absorbed by interstellar matter.
The High-Energy X-ray Timing Experiment \citep[HEXTE;][]{Rothschild}
was used to extend the bandpass with 
30--200 keV. 
This allowed us to acquire bolometric correction factors (see G08) 
on the PCU2 spectra for six distinct epochs \citep[see also][]{Zhang09}, 
in order to 
estimate
the bolometric persistent fluxes 
between 0.1--200 keV.
 
We searched for, and analysed the bursts detected by the PCA following G08. 
We computed 1-s light curves on the full PCA energy range (2--60~keV) using 
'Standard-1' mode data and searched for significant deviations from the mean level. 
For each candidate event we identified, we computed burst spectra using (typically) 
64-channel 'Event mode' 
data from before the burst start, through to late in the burst tail 
(typically $\approx200$~s following the burst start). 
The integration time for each spectrum was 0.25~s at the start of the burst, increasing through 
the tail as the burst flux decreased to maintain approximately the same signal to noise level. 
We used a 16-s persistent spectrum (including the instrumental background) extracted from before 
the burst start as background for the burst spectral analysis. We fit each spectrum initially 
with an absorbed black-body, with the neutral column density $N_{\rm H}$ free to vary, and calculated 
its mean value over the burst. 
Subsequently we re-fit the burst data with $N_{\rm H}$ frozen at the mean level and computed 
the bolometric flux and uncertainties based on the measured black-body temperature and radius. 
Fluxes were eventually corrected for the off-axis pointing by scaling the measured flux with 
the ratio of the response of the collimator at the aimpoint to its response at the source position.

\subsection{Burst properties}  
\label{subsec:properties} 
We apply the definitions by G08 for burst start and rise times, as the time when the burst flux 
first exceeds $25\%$ of the peak flux, and the following interval 
until the burst exceeds $90\%$ of the peak flux, respectively.
We define the burst duration as the interval from the start time to the time when 
the burst flux has decayed to 25\% of the peak flux.
All uncertainties 
are given at a 1$\sigma$ confidence level. 

We compute the burst fluence $F_b$ by integrating the measured flux over the burst 
duration; the burst timescale $\tau$ as the ratio of the fluence to the peak flux; 
and $\alpha= (F_{\rm pers}/F_{b})\Delta t_{\rm rec}$ 
is the energy ratio between persistent emission and bursts \citep[e.g.][]{lewin83}, 
where $\Delta t_{\rm rec}$ is the recurrence time between two consecutive bursts, 
$F_{\rm pers}$ is the persistent flux at the time of the $2^{nd}$ burst, and 
$F_{b}$ is the fluence of this burst.

\section{Results}  
\label{sec:results}
\subsection{Persistent emission}  
\label{subsec:persistent} 

We follow the changes of the broad-band persistent emission from \IGR\ 
by combining BAT daily averaged data with all available \R\ data 
between MJDs 54500 and 54750 (MJDs 54560--54740 for PCA and HEXTE). 

\begin{figure}
\centerline{\epsfig{file=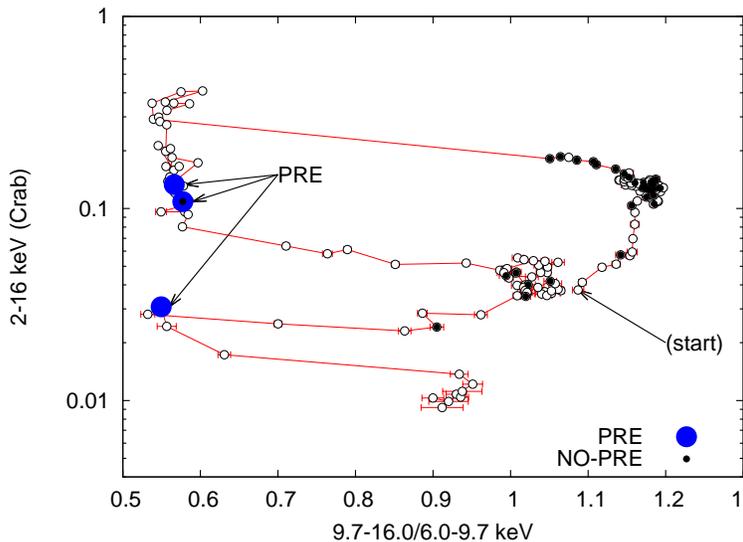,angle=-90,width=10.cm}} 
\caption 
{X-ray hardness-intensity diagram obtained from \R/PCA data for \IGR\ during 
its outburst in 2008. 
Hardness and intensity are normalised to the Crab. 
The positions of the X-ray bursts observed by \R\ are shown through filled circles 
(large for PRE bursts). The arrow indicates the first measurement at MJD 54560.} 
\label{fig:HID}
\end{figure}

Fig. \ref{fig:broad_outburst} depicts the evolution of the bolometric flux 
of \IGR\ together with ASM and BAT daily averaged light curves for the soft and hard 
persistent intensities, respectively.
The outburst started at the end of March, around 
MJD~54556 \citep[see also][]{Alta1459, Mark1460, kuul1461}, 
with a rapid rise in intensity, particularly in hard ($>15$~keV) X-rays, 
and reached a bolometric flux  $\simeq~1.3\times10^{-8}$~\ergcs~in less than 20 days. 
The source remained close to this flux level 
during the following one and an half month (from 
around MJD~54580 to 54625). 
The flux then rose for another ten days,
reaching a peak about $18.5\times10^{-9}$~\ergcs~on June 17 (MJD~54634). 
At the same time the 
persistent spectrum dramatically switched from hard 
(i.e. with more flux in the 15--50~keV band than in the $<12$~keV band) 
to soft in less than 3 days, 
as witnessed by ASM and BAT light curves (Fig. \ref{fig:broad_outburst}). 
These two states are also obvious in the HID 
plotted in Fig. \ref{fig:HID} 
that shows the transition with the 9.7--16/6--9.7 keV hardness falling abruptly 
from above 1 down to below 0.6 at a 2--16 keV flux slightly increasing above 0.2 Crab.
The bolometric flux subsequently decreased almost continuously for about one month 
(from MJD~54635 to 54670), down to $\simeq 3\times10^{-9}$~\ergcs. 
From MJD~54670.5 the source spectral state returned from soft to hard at relatively low level 
(corresponding to the group near the middle of the HID at the hardness around 1),
remaining there for a month at a bolometric flux level about $4\times10^{-9}$~\ergcs, 
and finally decreased, with a hardness zigzag path, 
back to its pre-outburst level after another month \citep[see also][]{Zhang09}.

We note that the HID shows some similarities with corresponding diagrams
for black-hole X-ray binaries (BHXRB; see, e.g., Fender et al. 2004), 
and similarities in this sense have already been reported
(e.g. Fender \& Kuulkers 2001; Fender et al. 2003; Tudose et al.,2009).

\subsection {Burst analysis}
\label{subsec:bursts}
We found a total of 57 individual X-ray bursts from \IGR\ in 2008.
One of these bursts, on April 8 (MJD~54564.567), was simultaneously detected by 
\I/JEM-X and \R/PCA.
We have performed separate light curve and spectral analyses of each burst data 
set from the different instruments: SuperAGILE: 1 burst, XRT: 1 burst, 
JEM-X: 14 bursts (of which 6 were also weakly detected by ISGRI), 
and PCA: 42 bursts.

\subsubsection{Burst light curves} 
\label{ssubsec:lcburst} 
The first X-ray burst from \IGR\ in 2008 was detected by SuperAGILE on March 26 
(MJD~54551.972) at a time when the source was in relative quiescence 
at a 2-10 keV flux below $\simeq~10^{-10}$~\ergcs \citep{DelMonte, Mark1460, kuul1461}.
The event was visible only in the SuperAGILE 17--25 keV band light curve 
(see Fig. \ref{fig:SA_LC}) for a duration of 44~s. 
The broad single peak, showing some hints of internal structure, was not preceded by a 
precursor. 
From the SuperAGILE image in the same energy band, the average flux is
derived to 
$\simeq 3.1\pm 0.6 \times 10^{-9}$~\ergcs~and the burst fluence is 
$\simeq 1.4\pm 0.3 \times 10^{-7}$~\ergcm. 
The hardness ratio (counts between 20 and 25 keV divided by the counts between 17 and 
20 keV) does not show any evolution during the burst. 
We did not find other bursts from the same source in SuperAGILE data during the whole 
AGILE observation from 20.7 days before to 3.5 days after the burst.

\begin{figure} 
\centerline{\epsfig{file=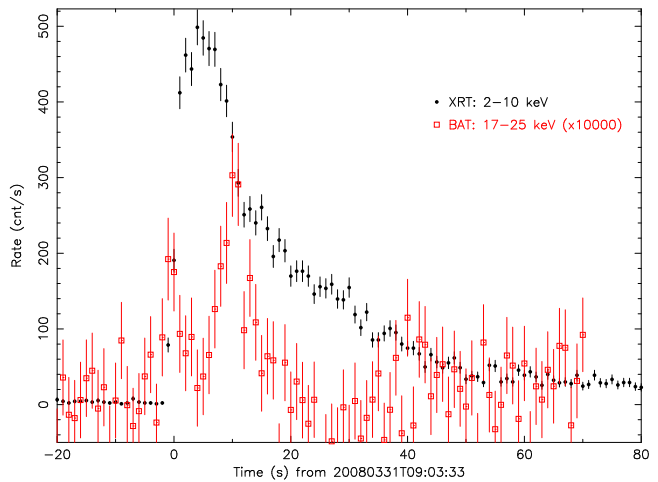,angle=-90,width=8.5cm}} 
\caption 
{1~s bin light curves of the X-ray burst observed by \S\ on March 31 (MJD~54556.377).} 
\label{fig:Swift_LC}
\end{figure}

A few days later, on March 31 (MJD~54556.377), the XRT and BAT instruments 
detected a second X-ray burst (see Fig. \ref{fig:Swift_LC}) 
when the 2--10 keV source persistent flux was still 
low at $\simeq 3\times10^{-10}$~\ergcs \citep{Alta1459}.
The 2--10 keV burst rise time is 4.5~s and the 
bolometric 
peak flux is $1.09\pm 0.08 \times 10^{-7}$~\ergcs. 
The decay consists of a first prompt decrease with an e-folding decay time of 12~s, 
followed 14~s after the burst peak by a longer tail decreasing 
with an e-folding decay time of 25~s; the single exponential decay time for the whole light curve is 21~s.
The 17--25~keV BAT light curve (see Fig. \ref{fig:Swift_LC}) 
allows us to compare this burst with the SuperAGILE one 
in the same energy band, and indicates a burst duration 
$\simeq~20$~s, about a factor of two shorter than the burst observed by SuperAGILE. 
In this energy band the average BAT burst flux is $\simeq 3.8\pm 1.5 \times 10^{-9}$~\ergcs~and 
the fluence is about $0.7\pm 0.1 \times 10^{-7}$~\ergcm. 
Moreover, we note that the hard X-ray light curve 
shows a double peaked structure, which is often related to a PRE event 
(see section \ref{ssubsec:burstspec}). 

\begin{figure*} 
\centerline{\epsfig{file=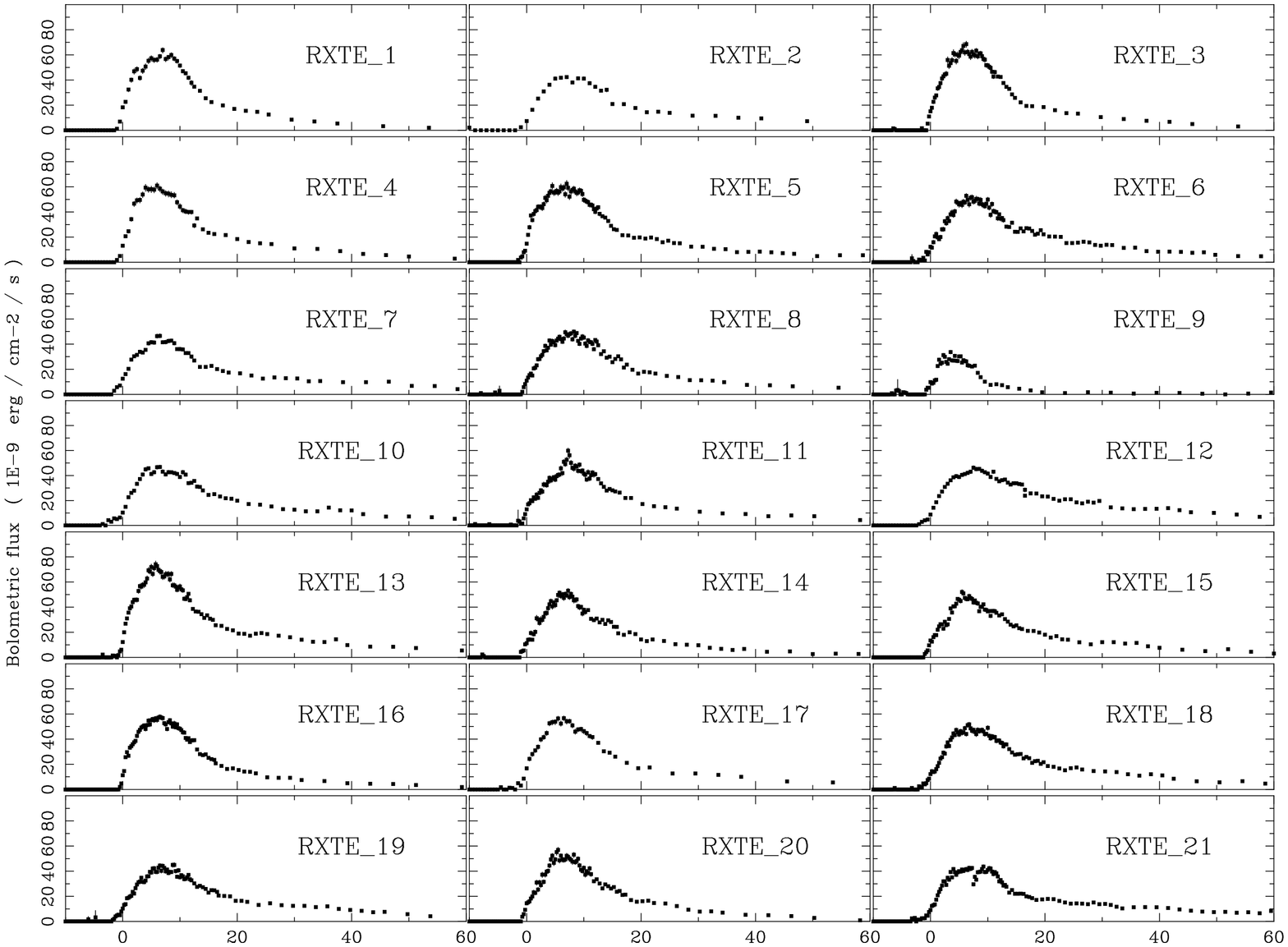,angle=-90,width=15cm}}  
\centerline{\epsfig{file=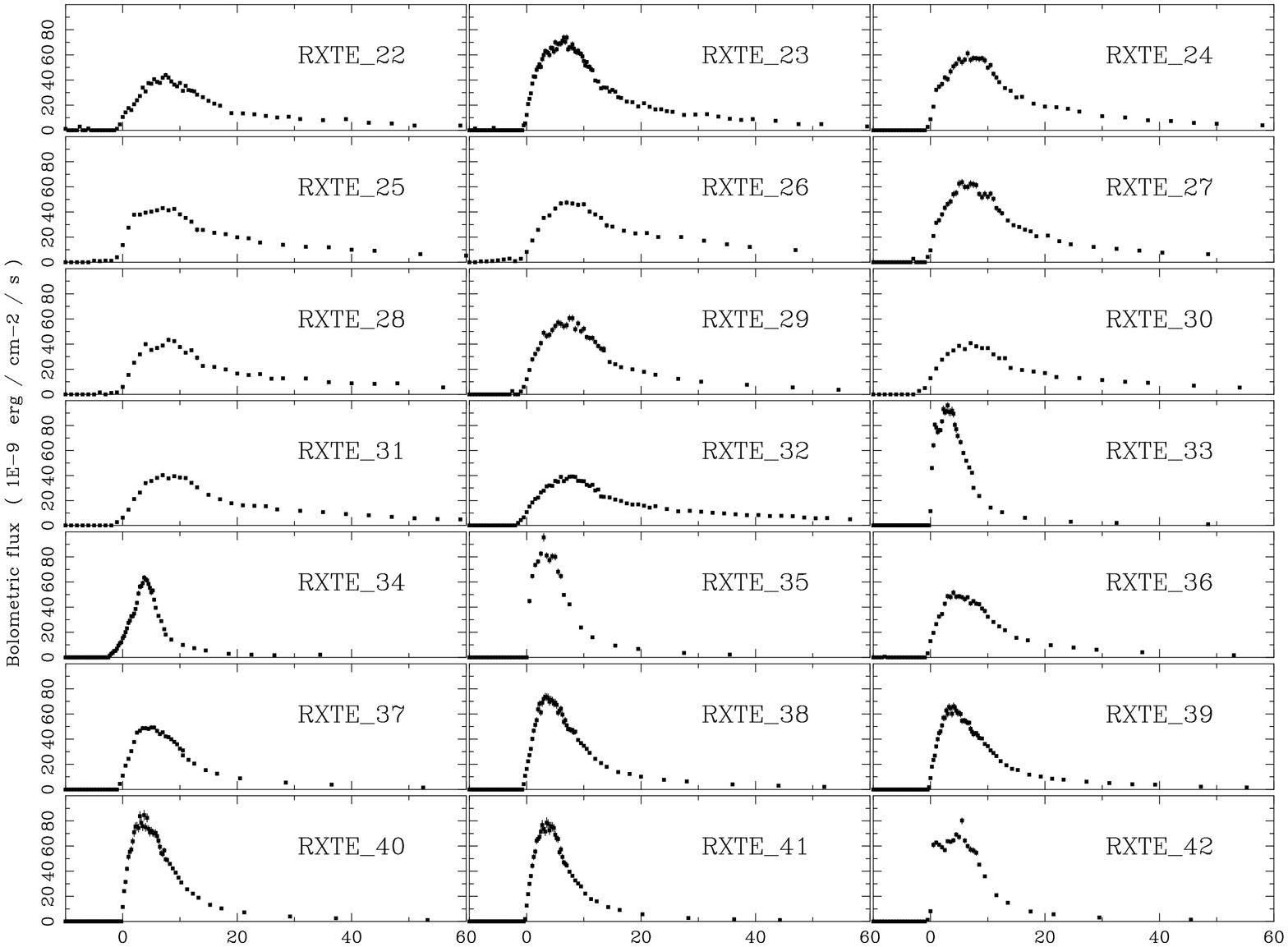,angle=-90,width=15cm}} 
\caption 
{Chronologically ordered bolometric flux light curves of the X-ray bursts observed by \R/PCA.
}
\label{fig:PCA_LCb}
\end{figure*}

\I\ detected 14 X-ray bursts near the beginning and the end of the outburst of \IGR\ 
(see Fig. \ref{fig:JMX_LC} and Table \ref{tab:burst_lst}). 
Thanks to the long uninterrupted \I\ observations on the source, we recorded five pairs of bursts
from which we can unambiguously measure the wait times between the bursts, which
decreased from 6.5~h on MJD~54562 down to 2.1~h on MJD~54576.
The first 11 bursts, observed during the outburst rise, have relatively long rise times 
of between 3~s and 9~s ($5.2 \pm 1.8$~s on average) and 
long exponential decay times of between 10~s and 33~s ($14.5 \pm 4.5$~s on average). 
The remaining three bursts, observed during the outburst tail,  
evolved more rapidly, with rise times of between 1~s and 5~s 
(average $=2.7 \pm 1.0$~s), and decays of between 9~s and 14~s (average
$10.3 \pm 1.4$~s). 
Among the bursts also detected by the IBIS/ISGRI instrument, two were significant up to 40 keV. 
As a comparison with the two first bursts observed by SuperAGILE and \S, 
the 17-25 keV light curve of the first burst observed by \I\ on April 6 (MJD~54562.508) 
indicates a duration of only $\simeq~8$~s.

The richest data set is provided by PCA, 
covering most of the outburst. 
The bolometric flux light curves of the 42 bursts detected by PCA have been 
compiled in Fig. \ref{fig:PCA_LCb} (see also Table \ref{tab:burst_lst}).
Burst \#1 
is the event 
simultaneously detected by JEM-X on MJD~54564.567, 
though their light curves are not easy to compare because the JEM-X and PCA 
instruments have different responses.
Bursts \#8 and \#9 form a doublet only separated by 510~s.  
Observations of a few burst pairs indicate a steadily decreasing wait time 
to a minimum of 1.6~h at MJD~54585.
We notice a prolonged pause of burst activity 
after the outburst emission reached its bolometric peak flux, also corresponding to the sudden 
change in spectral hardness (Figs. \ref{fig:broad_outburst} and \ref{fig:HID}). 
The last recorded burst before the peak of the outburst is \#32, 
and the next recorded burst, \#33, 
was observed one month later. 
The difference between the light curves of these two bursts is clear and 
indicates a major change in the general shape of the burst light curves before and after 
the outburst peak.
Thanks to the good time resolution of the PCA, the light curves can be fit with dual exponential decays.  
On average, the 32 PCA bursts that occurred before the peak of the outburst 
have rise times of $6.6 \pm 1.1$~s, and exponential decays of $9.7\pm2.7$~s and $26.8 \pm 7.4$~s. 
The 10 remaining bursts after the resumption of the burst activity during the decay of 
the outburst are characterised by much shorter rise times of $4 \pm 1$~s on average, 
and exponential decay times of $5.2 \pm 1.6$~s and $14.5\pm 4.2$~s on average.
This is consistent with the bursts reported by JEM-X at corresponding epochs.
Among the PCA bursts, \#33, which is the first burst observed after the one-month intermission, 
has the highest peak flux at $F_p=9.6 \pm 0.2 \times10^{-8}$~\ergcs, 
as well as the shortest rise time of $2.25 \pm 0.25$~s 
and shortest exponential decay time of $4.8 \pm 0.5$~s. 

\subsubsection{X-ray burst spectral analysis} 
\label{ssubsec:burstspec} 
\begin{figure} 
\centerline{\epsfig{file=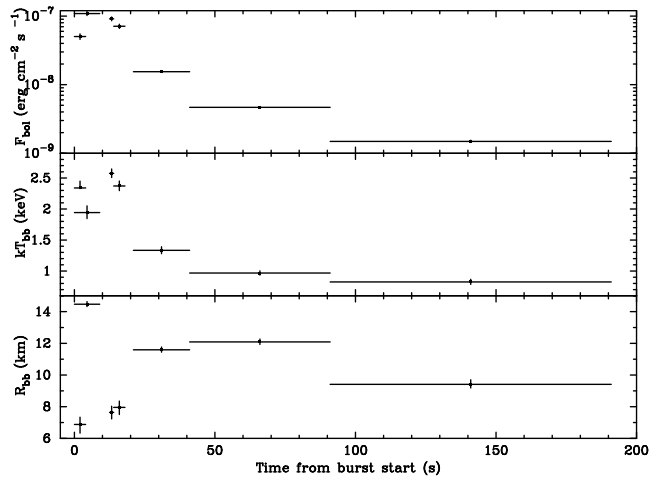,angle=-90,width=8.5cm}} 
\caption 
{Time-resolved spectroscopy results of the X-ray burst observed with \S\ on March 31 (MJD~54556.377). 
The black-body radius is calculated from the fit normalisation for a distance of 5.5~kpc 
(see section \ref{subsec:Summary}).} 
\label{fig:Swift_ana}
\end{figure}

We present results of time-resolved spectral analyses performed for the bursts observed 
with \S\ (see Fig. \ref{fig:Swift_ana}) and with \R/PCA (see Figs. \ref{fig:kTb} and \ref{fig:Radb}).

The time-resolved spectroscopy of the \S\ burst reveals anti-correlated black-body 
temperature and radius 
variations at roughly constant bolometric flux during the first 20 seconds.  
The PCA bursts \#33, \#35, and \#42 exhibit the same properties, characteristic of PRE events. 
The average bolometric peak flux of these four bursts is $F_{peak}=(95\pm9)\times10^{-9}$~\ergcs, and 
the highest peak flux is reached by the \S\ burst at $F_{peak}=109\pm9)\times10^{-9}$~\ergcs. 
Assuming that the highest peak flux 
corresponds to the Eddington luminosity limit for helium bursts, $L_{\rm Edd}=3.8\times10^{38}$\ergps, 
as empirically derived by \citealt{kuul03} (but see also section \ref{subsec:Summary}), 
we derive a source distance of $5.4 \pm 0.2$~kpc \citep[see also][]{Alta1459, Alta1651}.
In Figs. \ref{fig:Swift_ana} and \ref{fig:Radb} the black-body radius is presented 
at a distance of 5.5~kpc (see section \ref{subsec:Summary}). 
It is worth noting that all these PRE flashes occurred at times when the 15--50~keV intensity 
measured by the BAT was at its lowest level (see Fig. \ref{fig:broad_outburst}) during the 
soft spectral state.
This can also be seen in the HID of the outburst Fig. \ref{fig:HID}, 
placing the PRE bursts in the softest states.

We further notice that bursts \#38 to \#41, whose temperatures peak above 3~keV, 
show slightly anti-correlated variations of the black-body radius with the colour temperature, 
in the opposite way as PRE events. These bursts have also in common to all occur at 
low persistent bolometric flux $<3.5\times10^{-9}$~\ergcs.

Time-resolved spectral analysis results of the bursts observed by 
the other instruments do not provide additional information. 
The complete analysis results for every burst are compiled in Table \ref{tab:burst_lst}.

\subsection{Burst behaviour as a function of the bolometric flux} 
\label{subsec:burstacc}
During the rise phase of the outburst bolometric flux 
the burst recurrence time became progressively shorter, down to less than 2~h, 
as derived from uninterrupted observations of the source between pairs of consecutive bursts. 
The light curves of the bursts observed in that phase of the outburst showed 
slow rise and relatively long decay times (Figs. \ref{fig:PCA_LCb} and \ref{fig:JMX_LC}).
After the persistent flux reached its maximum, corresponding to $\simeq 15-20\%$ 
of the Eddington limit (see Fig. \ref{fig:broad_outburst}), 
no bursts were detected during a prolonged period of low flux in a soft
spectral state.
The burst activity resumed after an interruption of one month,
with shorter and more intense bursts than before the peak of the outburst.
Figure \ref{fig:tau_vs_pers} shows a likely correlation (Pearson's coefficient r=0.62) 
between increasing exponential decay times and increasing persistent flux. 
On the other hand, we also notice that the bursts with the highest peak fluxes 
tend to have the shortest decay times (see Fig. \ref{fig:peak_vs_tau}). 

Apart from the PRE bursts only occurring in the soft state as already mentioned, 
we also notice in the HID Fig. \ref{fig:HID}, that there are more bursts in the hard state 
than in the soft state, all above a 2--16~keV intensity of 0.25 Crab.
The bursting activity seems thus also related to the source spectral state.
 
\begin{figure} 
\centerline{\epsfig{file=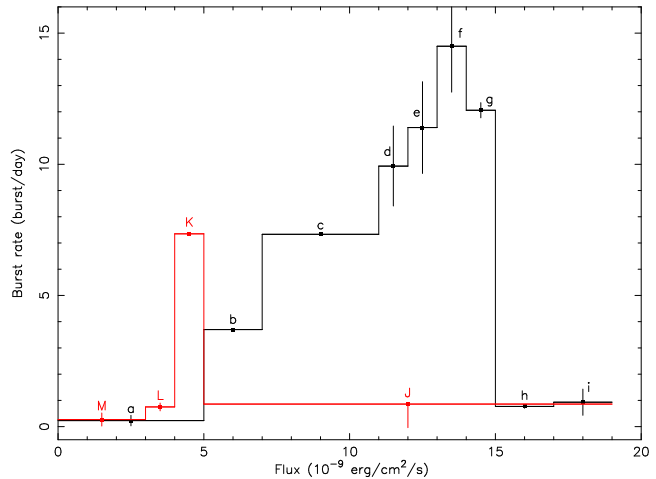,angle=-90,width=8.5cm}} 
\caption
{
Plot of the \IGR\ burst rate as a function of persistent bolometric flux 
derived from the waiting times between two observed consecutive bursts. 
Bins in black (lower case letters) and red (upper case letters) correspond to 
pre- and post-outburst peak, respectively.
The letters indicate the number of bursts ($n$) and exposure times ($exp$ in ks) per bin as follows: 
$n/exp$ --- 
a: 2/11.2, b: 3/39.8, c: 2/18.9, d: 8/50.6, e: 9/77.2, f: 8/44.1, g: 6/24.9, h: 5/16.8, 
i: 1/3.4, J: 0/60.2, K: 2/72.8, L: 9/55.1, and M: 2/40.9.
The heights of the bins without error bars are given by one pair of bursts detected during an 
uninterrupted observation. The error bars are obtained either by the difference 
between the waiting times of two distinct pairs of bursts observed inside the same flux interval, 
or by the difference between the averaged time from burst to burst and the shortest time 
within two consecutive burst detections in separated observations.
}
\label{fig:burst_rate}
\end{figure}

We adopt here the bolometric (0.1--200 keV) flux as a tracer of the accretion rate, 
but acknowledge that there may be uncertainties of order 40\% due to undetected soft components 
\citep[see][]{Thompson_Gal}, and  several per cents due to jet activity, though it is not yet clear 
how much it can be \citep{Russell}. 
The burst rate as a function of persistent emission (or accretion rate) 
plotted in Fig. \ref{fig:burst_rate} indicates a  
steady increase 
until a maximum of 14.5 bursts per day is reached 
at a bolometric flux of $\approx13.5\times10^{-9}$~\ergcs 
(the short doublet is not included in Fig. \ref{fig:burst_rate}). 
The bursting activity drops rapidly above a bolometric persistent flux of 
$\simeq15\times10^{-9}$~\ergcs, corresponding 
to  $\simeq 15\%$ of the Eddington accretion rate 
(see section \ref{ssubsec:burstspec}).
The red histogram (labeled with uppercase letters) of Fig. \ref{fig:burst_rate} displays the burst rate 
during the decrease of the persistent emission.
Indeed, no bursts at all were detected in the 30 days following the bolometric peak flux 
of the outburst, even though observations totalling approximately 80~ks
were made with {\it RXTE}. 
Before the burst interruption, JEM-X and PCA detected 42 bursts in a total of about 
280~ks exposure time, corresponding to an average burst rate of almost 13 bursts per day. 
Assuming the same burst rate, and that Poisson statistics describe the
number of bursts observed in any given observation, the probability of
detecting no bursts 
during the interruption
is $6.1\times10^{-6}$. Even at a burst rate four times lower, this probability remains below 5\%. 
Twenty eight \R\ observations, of 4.2~ks mean duration (with a standard deviation of 2.9~ks), 
were performed randomly during one month and separated on average by 1 day 
(15 hours standard deviation).
Therefore, we conclude the chance probability of having missed every burst in that period, 
to be negligible.
In other words, the burst rate must have dropped considerably below an upper limit 
that, from the available observations, we assess at $\simeq 0.86$ burst per day 
(at $1\sigma$), which corresponds to the label "J" in Fig. \ref{fig:burst_rate}. 
The burst activity thus resumed ("K") about one month after the outburst peak 
and decreased later on with the decreasing persistent emission ("L" and "M").
So, for the same range of bolometric flux, the bursting activity in \IGR\ strongly depends on 
whether the persistent emission is increasing or decreasing. 
Moreover, we note from Figs. \ref{fig:tau_vs_pers} and 
\ref{fig:burst_rate} that, 
apart from the two bursts observed by SuperAGILE and \S\ at the beginning of the outburst, 
the bursts observed during the rise phase of the outburst occurred
at a persistent bolometric flux above $5\times10^{-9}$~\ergcs, while all the bursts 
observed during the outburst decrease phase occurred below this threshold.

\section{Discussion}
\label{sec:Discussion}  
\subsection{Summary of main results}
\label{subsec:Summary} 

Thanks to the wide coverage by \R, \I, \S\ and {\it AGILE} and their 58 burst detections, 
the 6-month long outburst from the transient \IGR\ provides a comprehensive record 
of burst activity on a neutron star over a wide range of accretion rates.
The most interesting aspect of this record is the hysteresis shown in Fig. \ref{fig:burst_rate},
particularly the 1-month long intermission in burst activity that
commences with the peak of the accretion outburst and at the same time of 
a spectral state change in the accretion radiation. 
Similar burst intermissions also seem to have been exhibited by the bursting
transients EXO 1745-248 (G08) and 4U 1608-52 (Keek et al. 2008).

From the observation of PRE bursts we derive the distance to the source at $\simeq5.4$~kpc 
by assuming the highest burst peak flux equal to the Eddington limit for a H-poor photosphere: 
$L_{\rm Edd}=3.8\times10^{38}$~\ergps \citep{kuul03}.
It is however not possible a priori to know exactly the composition of the NS photosphere. 
The accreted material is most likely H-rich, based on the $\alpha$- values (see below), 
but the NS photosphere during radius expansion may be H-poor once the outermost, H-rich layers 
are ejected. Such a scenario is suggested from observations of bimodal radius-expansion burst peak  
fluxes in 4U 1636-536 \citep{sugi84, Gal06}, although the details of this mechanism remain uncertain.
Though the peak fluxes of the four PRE bursts are not exactly the same, we assume that they all 
intrinsically reach the same Eddington luminosity \citep[see][]{Gal03, Gal06}. 
Their characteristics (fast rise and short decay) being consistent with the burning of H-poor material, 
we interpret their peak flux as actually reaching the Eddington limit for He, 
and the source distance thus derived being comprised between 5.4 and 6.3~kpc. 
If one assumes the theoretical value of $L_{\rm Edd}\simeq2.9\times10^{38}$~\ergps for He bursts 
on a canonical 1.4 solar mass and 10 km radius NS \citep[see, e.g.,][]{GalCum06}, 
the distance is reduced to between 4.7 and 5.5~kpc. 
In conclusion, we find the distance to \IGR\ at $d=5.5 \pm 0.8$~kpc. 

A rare observation of an X-ray burst prior to the beginning of the outburst \citep{DelMonte} 
is the signature of some accretion activity at a particularly low rate \citep{Kuul09}.  
We estimate the 17--25 keV fluence of the SuperAGILE burst to about $1.4\times 10^{-7}$~\ergcm, 
which is twice higher than the \S\ burst fluence in the same energy band. 
For a 3 keV black-body model, this leads to a bolometric fluence of about $10^{-6}$~\ergcm, 
which is similar to the fluence of the confirmed PRE bursts observed later in the outburst. 
This fluence implies that accretion started at least 3 days prior to the burst observed by 
SuperAGILE, or a week before the onset of the outburst. 
Alternatively, it means that some fuel was left over from the previous outburst in 2005. 

We found a burst doublet with less than 10 min. recurrence time. 
While the first of these two bursts shows a relatively long rise (6.5 $\pm$0.3~s) 
and decay ($>$10~s) times, the second one has much shorter rise (3.5 $\pm$ 0.3~s) and 
decay (5 $\pm$ 0.5~s) times. The fluence of this latter burst, equal to $3.49\times10^{-7}$ \ergcm, 
is 3 to 4 times lower than the previous and the other neighbour bursts. 
Moreover, assuming complete and isotropic burning of the fuel accreted 
between these two bursts, the ignition column for the $2^{nd}$ one is: 
$y = F_{\rm  pers}\times(d/R)^2\times (1+z)/(z~c^2) \times \Delta t\approx 10^7$~\gcm2. 
This small ignition column, about one order of magnitude lower than for typical bursts, 
is insufficient to ignite freshly accreted nuclear fuel. 
It may therefore indicate that the second burst is related to the first one, 
and produced by some residue of unburnt fuel as suggested by \citet{Fuji87}. 
However, other examples of burst doublets have already been observed from other sources 
with even shorter intervals \citep[see, e.g., G08,][]{Linares1979} and their explanation 
may be related to the mixing of the unburnt fuel in deeper layers \citep{Boirin07, Keek10}.

One convenient way to investigate the burst energetics is to compare the amount of energy 
radiated by the persistent emission between the bursts with the energy released by 
the thermonuclear bursts. 
Their ratio is the 
$\alpha$ parameter, determined observationally as the ratio of the integrated persistent (bolometric)
flux over the burst interval to the burst fluence (see section \ref {subsec:properties}).
Assuming that all the accreted fuel is burnt during the bursts, 
the nuclear energy release per nucleon
\citep[see, e.g.,][]{GalCum06} is: 
$Q_{\rm nuc} = z~c^2/\alpha\, (10^{18}\,$erg/g)$^{-1}$, 
where $z=0.31$ is the appropriate gravitational redshift 
at the surface of a 1.4 M$_{\odot}$ and 10 km radius NS.
For a given average mass fraction $X$ of H at ignition, one has: 
$Q_{\rm nuc} = 1.6 + 4X$ MeV nucleon$^{-1}$.
During the plateau prior to the peak of the outburst of \IGR, $\alpha$ is about 75 -- 80, 
increasing to $\alpha~\ga110$ when the bursting activity resumes after the 
peak of the outburst.
These values correspond to H fractions in the burst fuel of
$X\approx 0.50$ and
$X~<0.23$, respectively.
This difference in fuel composition, and thus in burning regime, may explain 
why the burst rate shown in Fig. \ref{fig:burst_rate} is higher after the outburst peak 
than compared to when it was at the same flux level at the beginning of the outburst.

Eventually, the originality with \IGR\ is that 
seven phases of burst behaviour can be distinguished 
in Figs. \ref{fig:broad_outburst} and \ref{fig:PCA_LCb}: 
\begin{enumerate}
\item Very early in the outburst, a first burst detected with SuperAGILE, 
  followed a few days later by the \S\ observation of a PRE burst. 
  Both bursts occur at a low accretion rate $\lesssim$1\% of the Eddington limit;
\item During the rise of the outburst and the subsequent plateau
  phase in hard state (from the first JEM-X burst to \R\ \#24), 
  bursts arrive faster with increasing accretion rates. The
  slope of the linear rise in the burst rate versus persistent flux
  histogram (Fig. \ref{fig:burst_rate}) is equivalent to $\alpha=80\pm16$,
  which is consistent with the direct (burst-to-burst) measurements in
  Table \ref{tab:burst_lst}. Together with this value, the relatively long burst rises
  and decays point to He-ignited H/He bursts in which the H is
  burnt through the rp process \citep[compare with burst profiles of GS~1826-24,][]{Heger07};
\item During the final $\simeq$15 days leading to the outburst peak (\R\ bursts \#25 to \#32),
  when the bolometric flux is in excess of
  $15\times10^{-9}$~erg~s$^{-1}$cm$^{-2}$, the recurrence time drops
  by an order of magnitude (Fig. \ref{fig:burst_rate}). 
  This leads to an increase of alpha about 1500, 
  consistent with simultaneous stable and unstable He burning \citep{intZ03};
\item At the time of the outburst peak, a 1-month intermission in
  burst activity starts. The outburst peak also coincides with a sharp
  spectral transition into a soft state;
\item After the intermission, when the outburst is still in the soft
  state, but when the persistent flux is at similar levels as during
  the first few bursts early in the outburst, the brightest and
  shortest X-ray bursts occur (\R\ bursts \#33--35, two with PRE). Most
  likely these are pure He flashes;
\item After the persistent spectrum switches back to a hard state
  within 10~d, the X-ray bursts become longer again, but not as long
  as earlier in the outburst, and are probably again due to mixed He/H
  burning (JEM-X bursts \#11--14 and \R\ bursts \#36--41);
\item the last observed burst (\R\ \#42) occurs after yet another transition of the
  accretion radiation from a hard to a soft state at low accretion rate. 
  This burst also shows limited PRE.
\end{enumerate}

A study of the burst activity in \IGR\ was recently published in \citet{Chen2010}. 
As a main result, these authors find two parallel evolution groups of 
the burst durations as function of the persistent luminosity. However, we note that they only 
employ 16 bursts during the outburst of \IGR\ in 2008, all after MJD~54624, against 57 in our 
data set, which spans the whole outburst. We mean that the findings of \citet{Chen2010} derive 
from a selection effect, for no such parallel groups are visible in Fig. \ref{fig:tau_vs_pers}, 
which corresponds to Figure 5 of \citet{Chen2010}.
Moreover, we note a discrepancy between our bolometric fluxes and those of \citet{Chen2010}, 
which is most probably due to the limited energy range (1.5--30 keV) these authors adopt to define the 
bolometric luminosity based only on PCA measurements. 
Such a limited bolometric range does not adequately represent the persistent 
luminosity of \IGR, and hence its mass accretion rate during the different phases of the outburst.
Therefore, we do not come to the same conclusions as \citet{Chen2010}, but rather suggest  
the following interpretations of our observation results.

\subsection{Possible interpretations}
\label{subsec:Interpretations}

There is considerable variability in burst peak fluxes and time
scales. The rise and decay times tend to be somewhat longer before the
intermission than after, particularly the rise times. The shortest
rise times combine with the highest peak fluxes. These trends indicate
a varying He to H abundance ratio in the burning
layer. This is not unexpected considering the varying accretion
rate. The shortest bursts (\R\ bursts \#33--35) are consistent with
pure He burning. All other bursts are longer and fainter and must
have larger H fractions in the fuel. Therefore, the mass donor
must be H-rich. The fact that pure He bursts occur implies
that stable H burning (through the hot CNO cycle) must
simultaneously be present during \R\ bursts \#33--35, particularly
since bursts at similar persistent fluxes early in the outburst (for
instance \R\ burst \#1) have roughly twice longer times scales and
half the peak flux. Thus, we identify four burst regimes. 
First, mixed H/He flashes during burst phases i,ii, iii and vi. 
Second, pure He flashes concurrent with stable H burning in burst phase v. 
Thirdly, rich He flashes at the lowest persistent fluxes. 
Lastly, no bursts at all and, therefore, stable H and He burning during the intermission. 
Thus, all burst regimes for H/He burning \citep{Fuji81, B98} are exhibited by \IGR, 
but at accretion rates about one magnitude order higher. 
Such a discrepancy between observed and theoretical accretion rates has already been 
noted for other sources \citep[see, e.g., the discussion by][]{cor03}.

The most intriguing question is what 
actually caused the 1-month interruption of
burst activity {\em after} the peak of the outburst. 
If the He and H are not burnt in an unstable fashion through flashes, they
must be burnt in a stable fashion. The drop in burst rate (Fig. \ref{fig:burst_rate}) 
just before the peak thus suggests the onset of stable H and He burning, at a
threshold persistent luminosity that is similar (i.e., within a factor
of two) to what is seen in other bursters (cf., Cornelisse et al. 2003), 
although the burst profiles in \IGR\ do not show as strong a
change. Hence, stable He burning, which should occur 
at higher accretion rates than stable H burning, apparently starts for \IGR\ 
at lower accretion rates than normally expected.
We discuss two scenarios.

One possibility is that the intermission is the result of thermal relaxation
of the crust.  The crust becomes hotter during the outburst due to
electron capture reactions (Gupta et al. 2007). The thermal time scale
of the crust is of order a few months (Brown \& Cumming 2009). It is
likely that the temperature is still rising after the outburst peak.
Above certain temperatures the H and He burning will become stable
\citep[$7\times10^7$ and $4\times10^8$~K, respectively; e.g.,][]{CM04}.
The intermission of bursts at the time of the peak of the outburst may
be explained if the temperature rises above the threshold temperature
for stable He burning just at that time. The thermal relaxation of
the crust may sustain that temperature for the duration of the
intermission after which it would drop below the threshold due to the
decay in accretion rate that started at the peak of the outburst.
Calculations for the transient 4U 1608-52, which had a similar
outburst duration and peak accretion rate as \IGR, indicate that the
temperature may reach high enough values (Keek et al. 2008). A problem
of this scenario is that the heating time scale appears short if the
burning depth is the same for stable H and stable He burning. 
Indeed, it takes for \IGR\ less than 15 days from the onset of stable H burning (drop of burst rate) 
to reach stable-only He burning (burst interruption), 
corresponding to a difference factor of 5--6 in temperature. 
The rapidity of this heating may be explained by the onset of stable H burning,
but verifying this requires solving the heat balance of the upper
layers. As with the calculation of the thermal relaxation of the
crust, this is outside the scope of this observational paper.

An alternative and perhaps more likely explanation for the
intermission is that a superburst occurred during a data gap, 
and provided the sudden heating to initiate stable He burning. 
There is a data gap of more than 10 hours between the last detected burst 
and the first available observation (by \R/ASM) of the soft state. 
Though the probability that a superburst just occurred during the data gap 
is rather small --- $\sim0.1\%$ for a typical superburst recurrence time of 
$\sim 1$ per year \citep{KeekZand08} ---
this is, in principle, enough time for a superburst to ignite and cool off.
Cessation of burst activity after a superburst is commonly
observed (e.g., Kuulkers 2004). This 'quenching' of normal bursts is
explained by the additional heat flux coming from the underlying
cooling ashes of the superburst that stabilizes the H/He burning
layers (Cumming \& Bildsten 2001). Cumming \& Macbeth (2004) predicts
timescales for quenching of weeks. In the cases of Ser X-1 (Cornelisse
et al. 2002) and KS 1731-260 (Kuulkers et al. 2002), no bursts were
detected after a superburst during a period of about one
month. Another example, more similar to the present case, is a
superburst from the transient and H-rich burster 4U 1608-52
(Keek et al. 2008), 55 days after the onset of an outburst in
2005. The first normal burst was detected 100 days after the
superburst, although long data gaps may have prevented earlier burst detections. 
Considering that the sudden burst intermission of \IGR\ began after 
the outburst peak at MJD 54635, a superburst would thus possibly 
have taken place more than 75 days after the outburst onset. 
The fluence of the outburst until the peak is 0.08~\ergcm, 
which at the distance to the source translates to an energy 
of $3\times 10^{44}$ erg. 
This is similar to the value reported by Keek et al. (2008) before the superburst 
of 4U~1608-522. The average bolometric flux of the outburst prior to the
maximum is $12.3 \times 10^{-9}$~\ergcs~or about 12\%
of the Eddington limit, which corresponds to the level of accretion
rate typically measured for superbursts \citep[see, e.g.][]{KeekZand08}. 
For $\Delta$t = 75 days, this corresponds to an accumulated
column depth $y = ˙m\Delta t/(1 + z) = 1.1\times 10^{11}$~\gcm2~.
This is one order of magnitude below the typical column
depth for superbursts whose energy release is $10^{42}$ erg, but a
part of the necessary C fuel may have been accumulated on a much
longer timescale at low accretion rate. Still, this value is only half
the value inferred from the superburst light-curve of 4U 0614+091 (see
Kuulkers et al. 2009), and as discussed by these authors, such a
shallow column depth would require for ignition that the temperature
of the NS crust is above $6\times 10^8$ K. 
As previously mentioned, both stable and unstable He burning likely occured 
during the 15 days prior to the outburst peak of \IGR, thus increasing considerably 
the C production. 
It is also worth noting 
that Cornelisse et al. (2002) find that the low energy flux increased
significantly after the superburst of Ser X-1 as well as after the
superburst from 4U 1735-44 (Cornelisse et al. 2000), like in IGR
J17473-2721.

Burst regimes appear to be tightly connected to positions in the HID,
see Fig. \ref{fig:HID}. The last five phases of burst behaviour (see above)
correspond to five distinct regions in the HID. A straightforward
explanation is that both have a common driver: the accretion rate.
Changes of burst regimes could be connected to threshold values in the
(local) accretion rate and likewise could changes in spectral states
of the persistent radiation be. What is remarkable for \IGR\ is that
one change in burst regime occurs at the same time and/or 
at same accretion rate as one change in the spectral state, 
and this is more clear than for other sources (see also G08). 
If one refuses this as a coincidence, one
could change the logic: the spectral state change may be due to the
missed superburst or the onset of stable He fusion process. 
Again, it is interesting to compare with 4U~1608-52, for which 
\citet{Yu_vdKlis} suggest that nuclear burning on the NS surface 
leads to the radiative truncation of the inner accretion disk.

In conclusion, the burst properties of IGR J17473-2721 are consistent
with a NS accreting H-rich material at varying rates all along the
outburst episode. The bursting rate as a function of the accretion
rate displays an hysteresis on either side of the peak of the
outburst, coincident with the interruption of the burst activity. 
This hysteresis may be explained by the thermal response of the NS crust, 
or the occurrence of a superburst. Though the odds of having missed a
superburst inside a data gap of only 10 hours are not high, we
consider that explanation as more likely than thermal relaxation of the crust. 
The question is whether the occurrence of this undetected
superburst is related to the abrupt spectral transition the source
underwent short after the peak of the outburst.

\section*{Acknowledgements}
We are grateful to Andrew Cumming and Rudy Wijnands for fruitful discussions. 
JC acknowledges financial support from the Instrument centre for Danish Astronomy and ESA/PRODEX Nr. 90057. 
\S/BAT transient monitor results are provided by the \S/BAT team and 
\R/ASM results are provided by ASM/\R\ teams at MIT and at the \R\ SOF and GOF at NASA's GSFC. 
This work is partly based on observations with \I, an ESA project with
instruments and science data centre funded by ESA member states (especially the
PI countries: Denmark, France, Germany, Italy, Switzerland, Spain), Czech
Republic and Poland, and with the participation of Russia and the USA.

\clearpage

\appendix
\suppressfloats
\section{Online-only figures and table}

\begin{figure}
\centerline{\epsfig{file=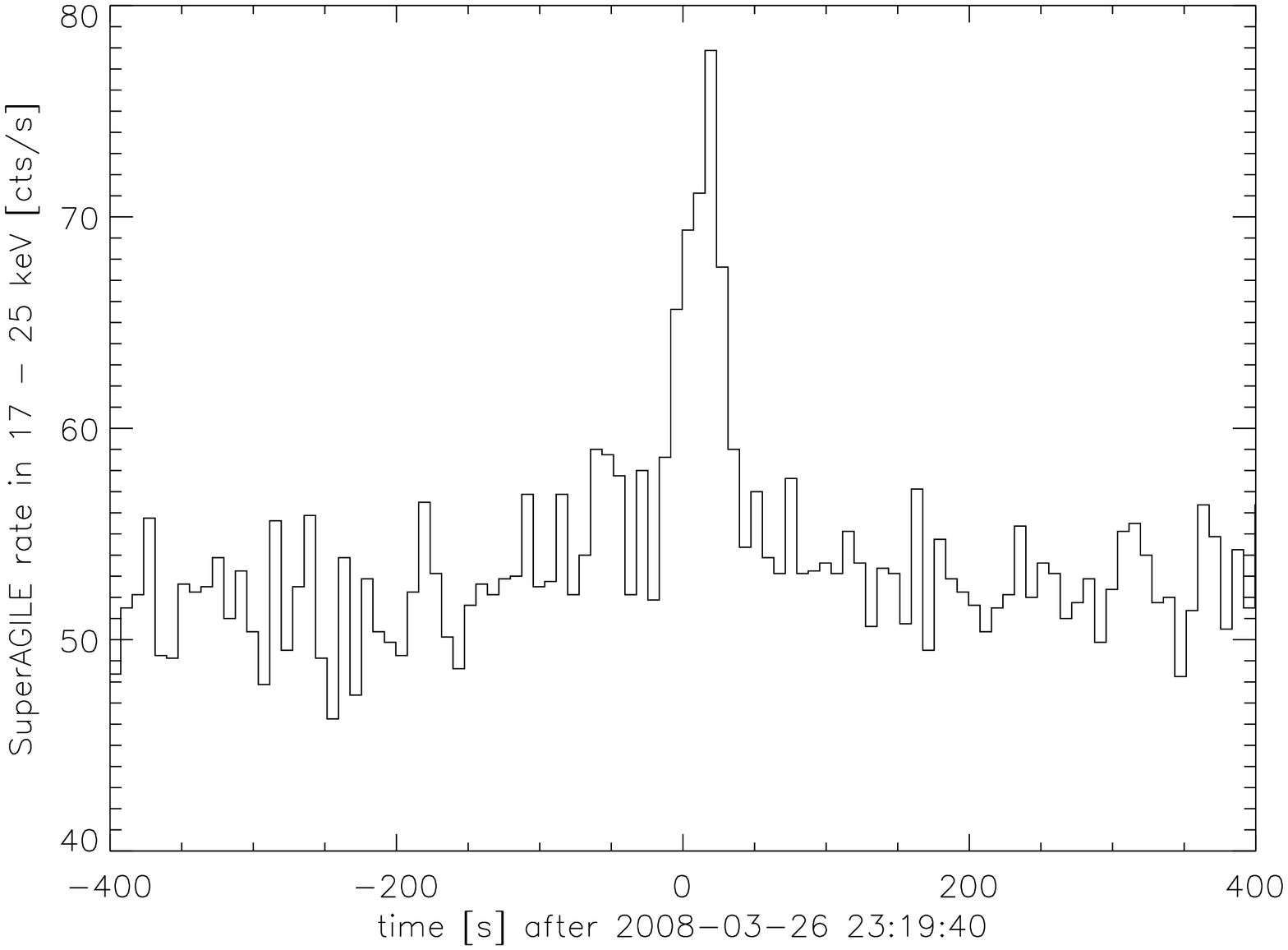,angle=90,width=8.5cm}} 
\caption 
{8 s bin, 17--25 keV light curve of the burst observed by SuperAGILE on 2008, March 26, 
(MJD~ 54551.972) prior to the onset of the \IGR\ outburst.} 
\label{fig:SA_LC}
\end{figure}

\begin{figure*} 
\centerline{\epsfig{file=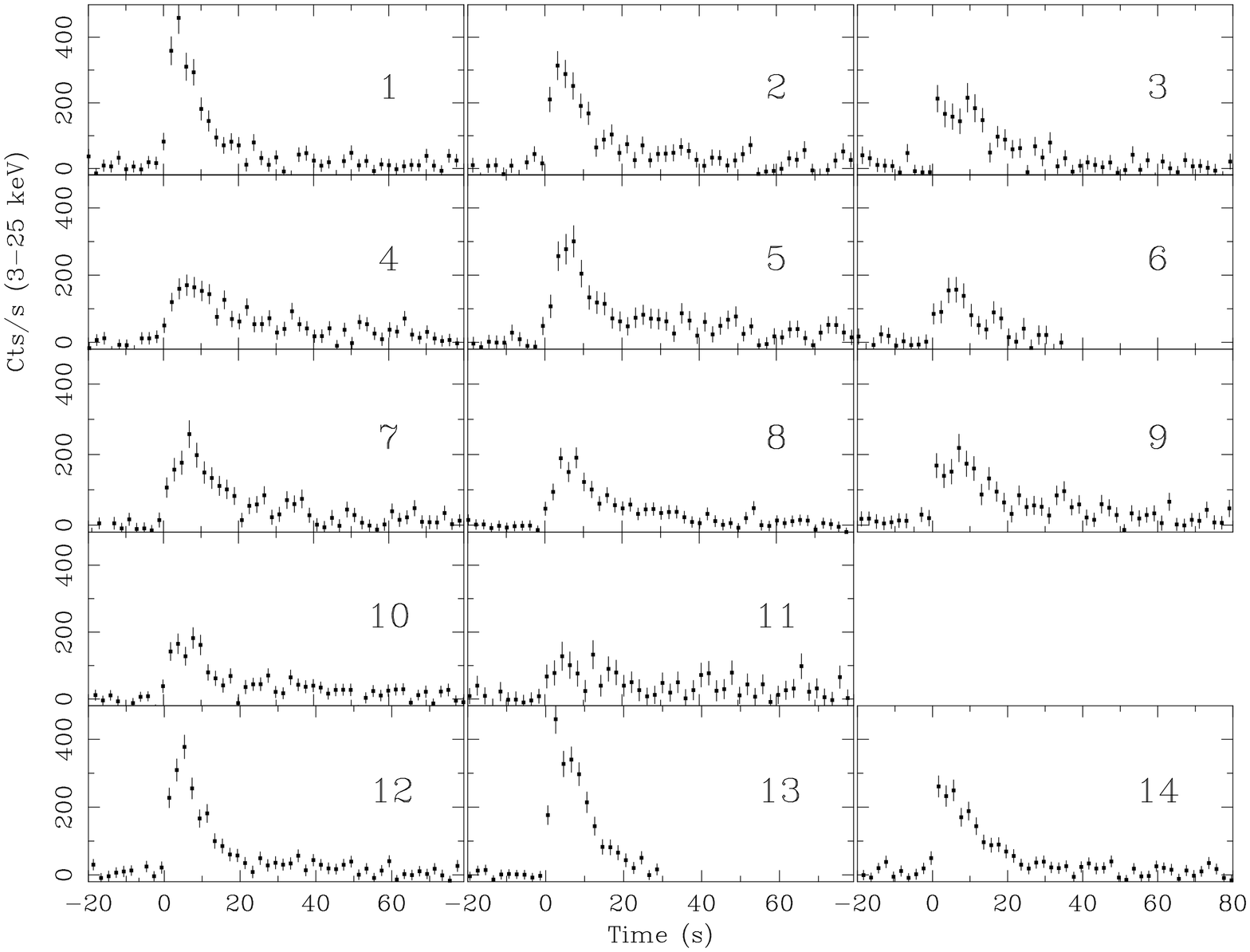,angle=-90,width=15cm}} 
\caption 
{Chronologically ordered 3--25 keV light curves of the X-ray bursts observed by \I/JEM-X 
between MJD 54562 and 54576 (from 1 to 11) and between MJD 54696 and 54711 (12-14). 
The vertical scale and the 2~s time binning are the same in every panel.} 
\label{fig:JMX_LC}
\end{figure*}

\begin{figure*} 
\centerline{\epsfig{file=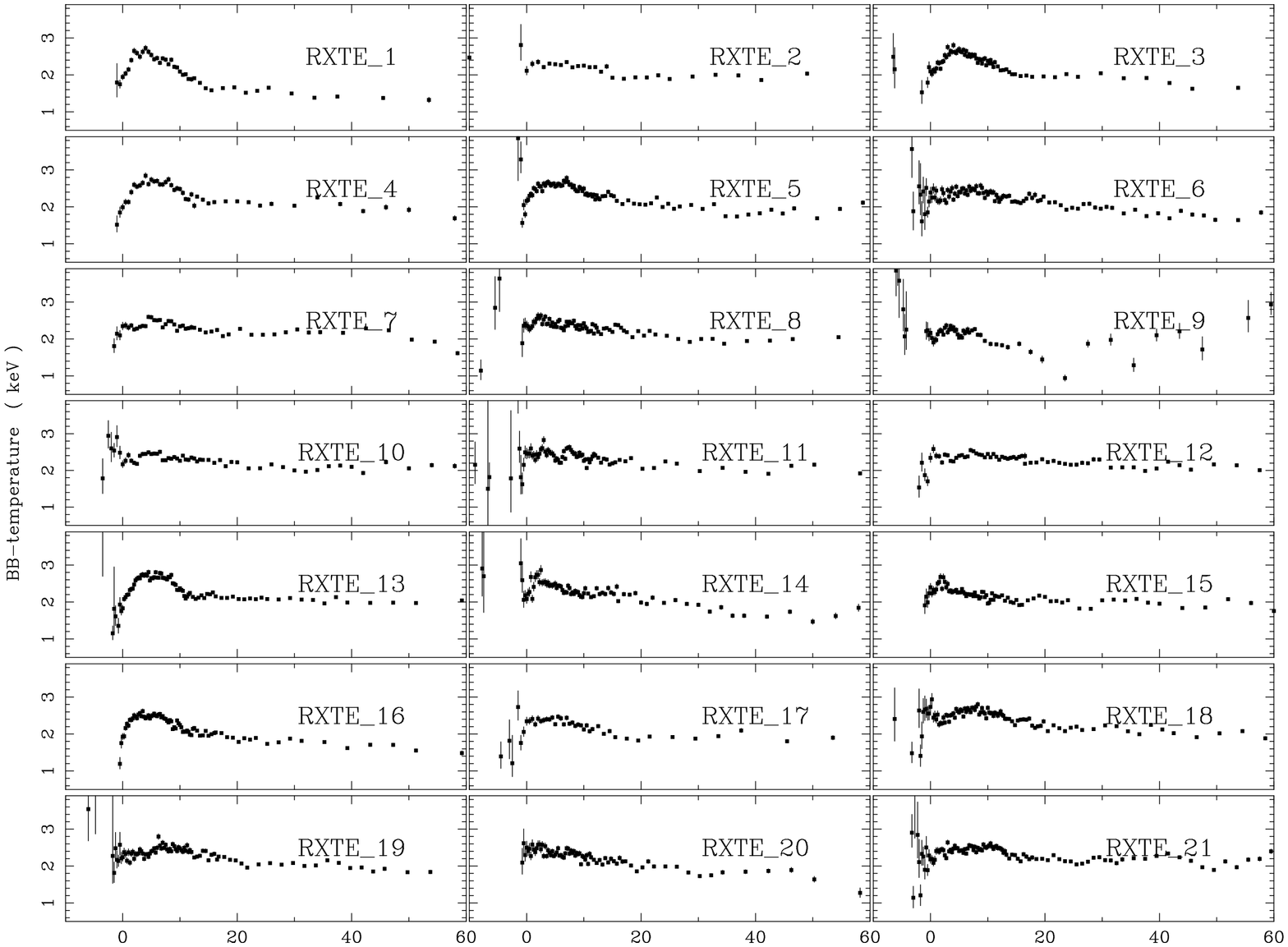,angle=-90,width=15cm}} 
\centerline{\epsfig{file=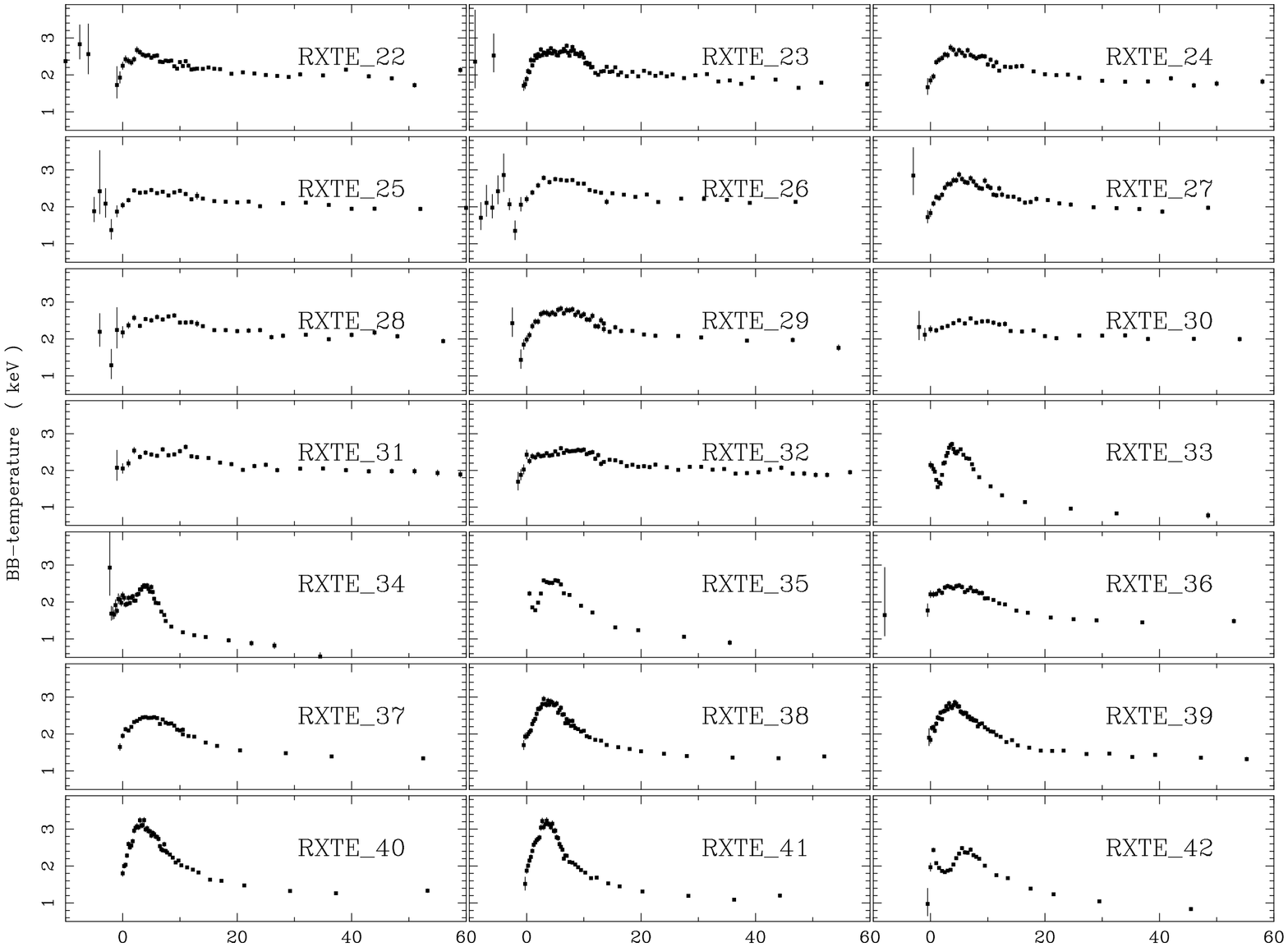,angle=-90,width=15cm}} 
\caption 
{Chronologically ordered kT variations of the X-ray bursts observed by \R/PCA.}
\label{fig:kTb}
\end{figure*}

\begin{figure*} 
\centerline{\epsfig{file=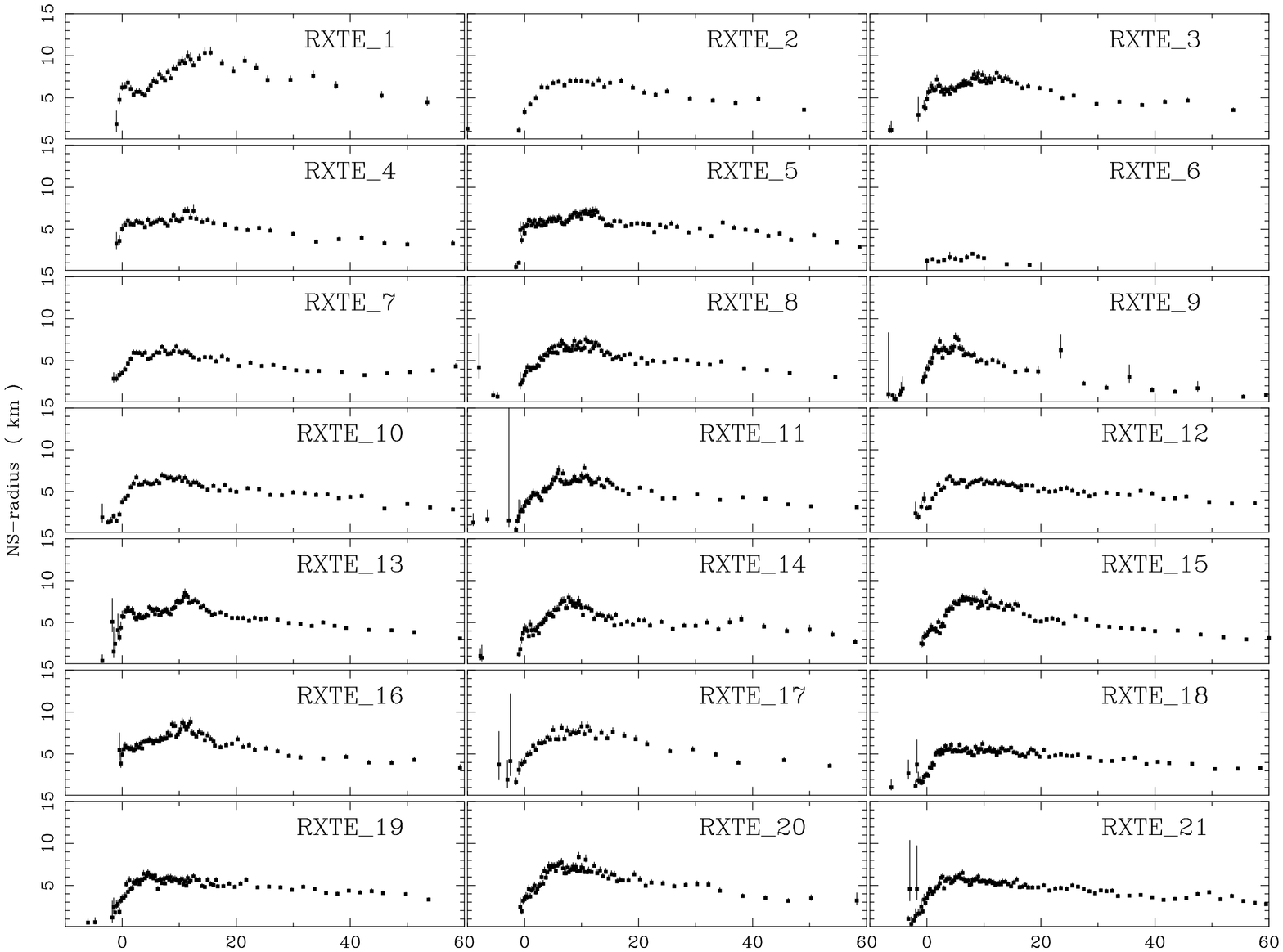,angle=-90,width=15cm}} 
\centerline{\epsfig{file=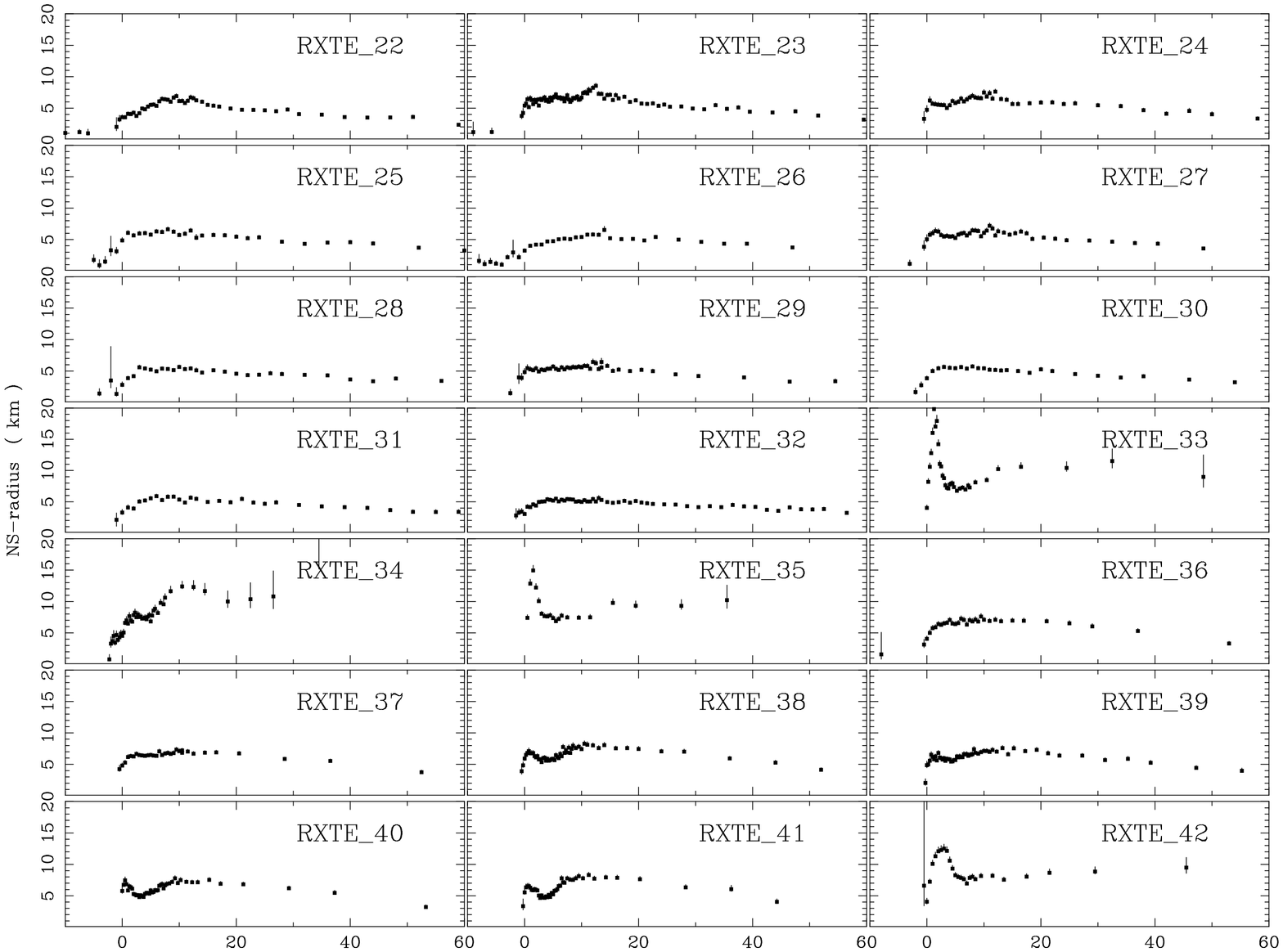,angle=-90,width=15cm}} 
\caption 
{Chronologically ordered black-body radius variations of the X-ray bursts observed 
by \R/PCA at a distance d=5.5~kpc.} 
\label{fig:Radb}
\end{figure*}

\clearpage

\begin{figure} 
\centerline{\epsfig{file=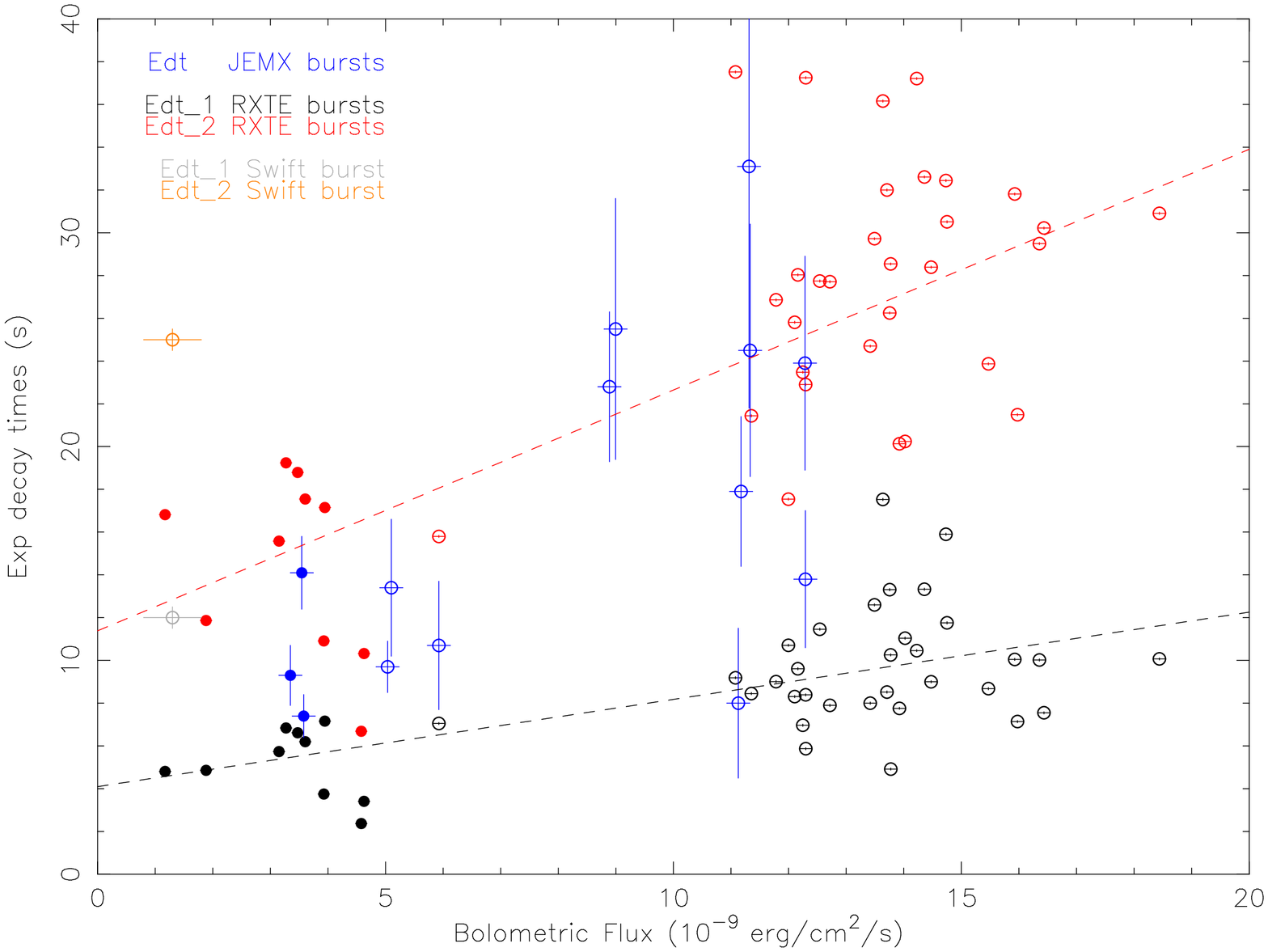,angle=-90,width=8.5cm}} 
\caption
{Plot of the \IGR\ bursts exponential decay times ("Edt") as a function of the persistent 
bolometric flux. 
Both \R/PCA (in black and in red) and \S/RXT burst (grey and orange) dual decays, 
as well as \I/JEM-X decay times (in blue) are shown. 
Open and close circles represent bursts observed before and after the peak of the outburst, 
respectively. The dashed lines show the linear regression fits of the PCA decays.
}
\label{fig:tau_vs_pers}
\end{figure}

\begin{figure} 
\centerline{\epsfig{file=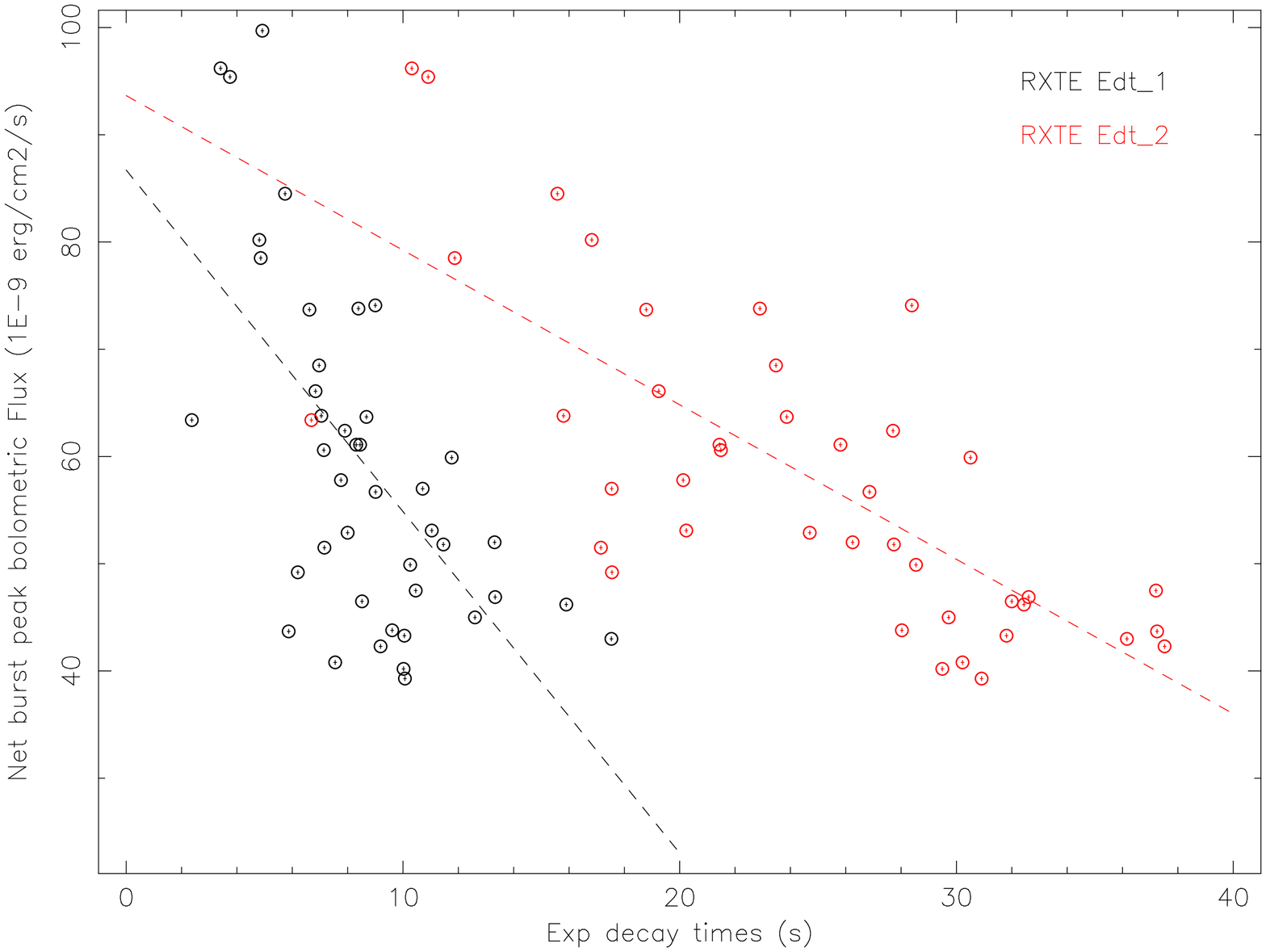,angle=-90,width=8.cm}} 
\caption
{Plot of the \R\ bursts peak fluxes as a function of the burst dual exponential decay times ("Edt").
The dashed lines show the linear regression fits.}
\label{fig:peak_vs_tau}
\end{figure}

\clearpage
 
\begin{table*}
\vbox to220mm{\vfil
\rotatebox{90}{
{\tiny
\begin{minipage}{236mm}
\caption{Summary of all burst observations from \IGR\ in 2008}
\label{tab:burst_lst}
\begin{tabular}{rccccccccccccc} 
\hline
 &  Date  &  MJD & PRE & $t_{\rm rise}$ $^{a}$  &  Edt1$^{b}$ &  Edt2$^{b}$  & $\tau^{c}$ & Duration & $F_{\rm peak}$ & Fluence & $F_{\rm pers}$ & $\alpha^{d}$ & $\gamma^{e}$ \\
 &  &  &  &  s  &  s  &  s  &  s  &  s  &  erg/cm$^2$/s  &  erg/cm$^2$  &  erg/cm$^2$/s  &  &  \\
\hline
AGILE  &  20080326T23:19:40  &  54551.972  & ? &  $\approx20$  &  &  & $\approx40$ &  44  &  $\approx3.5\times10^{-9}$  & (140$\pm$30)$\times10^{-9}$ &  $\sim10^{-10}$~?  &   & $\lesssim 10^{-3}$ \\
   (17-25 keV) \\
Swift &  20080331T09:03:33  & 54556.377  & \textbf{Y} &  4.5$\pm0.5$ & 12$\pm1$ &  25$\pm1$  & 16.8$\pm0.8$ & 30 &  ($\mathbf{10.9\pm0.1)\times10^{-8}}$  & (1828$\pm$70)$\times10^{-9}$ &  $\sim10^{-9}$  &   &  $3\times10^{-3}$ \\
   (2-10 keV) \\
RXTE  \\ 
   (Bolometric) &   &  &  &   $\pm$.25  &   $\pm$0.2  &   $\pm$0.2  &    &   $\pm$0.5  &  $10^{-8}$  &  $10^{-9}$   &  $10^{-9}$  &  &  $10^{-2}$ \\
1*  &  20080408T13:36:27  &  54564.567 &  N & 7.00  &  7.0  &  15.8  &  17.4 $\pm$1.2  &  21.5  &  6.4 $\pm$0.4  &  1111 $\pm$7  &  5.9 $\pm$0.6   &  --  &  5.4 $\pm$0.6 \\
2  &  20080414T07:38:24  &  54570.318 &  N & 6.00   &  9.2  &  37.5  &  30.6 $\pm$7.2  &  37.0  &  4.2 $\pm$0.4  &  1284 $\pm$179  &  11.1 $\pm$1.1  &  --  & 10.2 $\pm$1.1 \\
3  &  20080418T05:51:38  &  54574.244 &  N & 6.25   &  7.0  &  23.5  &  18.4 $\pm$1.7  &  23.8  &  6.8 $\pm$0.5  &  1251 $\pm$24  &  12.2 $\pm$1.2  &  --  & 11.2 $\pm$1.2 \\
4  &  20080420T08:24:32  &  54576.350 &  N & 6.00   &  8.4  &  21.4  &  20.2 $\pm$1.9  &  26.0  &  6.1 $\pm$0.4  &  1235 $\pm$33  &  11.3 $\pm$1.1  &  --  & 10.4 $\pm$1.1 \\
5  &  20080423T21:00:59  &  54579.876 &  N & 4.50   &  7.9  &  27.7  &  22.1 $\pm$2.3  &  28.8  &  6.2 $\pm$0.4  &  1372 $\pm$53  &  12.7 $\pm$1.2  &  --  & 11.7 $\pm$1.2 \\
6  &  20080424T20:22:18  &  54580.849 &  N & 6.25   &  8.0  &  24.7  &  23.8 $\pm$2.6  &  33.8  &  5.3 $\pm$0.4  &  1260 $\pm$43  &  13.4 $\pm$1.3  &  --  & 12.3 $\pm$1.3 \\
7  &  20080425T04:37:51  &  54581.193 &  N & 6.50   &  8.5  &  32.0  &  25.3 $\pm$3.7  &  32.5  &  4.6 $\pm$0.4  &  1164 $\pm$69  &  13.7 $\pm$1.4  &  --  & 12.6 $\pm$1.4 \\
8  &  20080426T00:50:50  &  54582.035 &  N & 6.75   &  10.3  &  28.5  &  24.0 $\pm$2.8  &  30.5  &  5.0 $\pm$0.4  & 1198 $\pm$46  &  13.8 $\pm$1.4  &  --  & 12.7 $\pm$1.4 \\
9  &  20080426T00:59:20  &  54582.041 &  N & 3.50   &  4.9  &  0.0  &  10.3 $\pm$1.5  & 11.5 & 3.4 $\pm$0.3 & 349 $\pm$20 & 13.8 $\pm$1.4 & 20.5 $\pm$1.0 & 12.7 $\pm$1.4 \\
10  & 20080426T23:59:39 & 54583.000 &  N & 6.00   &  13.3  &  32.6  &  27.6 $\pm$4.8  &  40.0  &  4.7 $\pm$0.4  &  1296 $\pm$116  &  14.4 $\pm$1.4  &  --  & 13.2 $\pm$1.4 \\
11  & 20080427T23:35:40 & 54583.983 &  N & 7.20   &  11.8  &  30.5  &  20.5 $\pm$2.8  &  26.2  &  6.0 $\pm$0.4  &  1228 $\pm$84  &  14.7 $\pm$1.5  &  --  & 13.5 $\pm$1.5 \\
12  & 20080428T01:37:29 & 54584.068 & N & 7.50   &  15.9  &  32.4  &  32.1 $\pm$5.9  &  49.5 & 4.6 $\pm$0.4 & 1478 $\pm$143 & 14.7 $\pm$1.5 & 73.2 $\pm$8.2 & 13.5 $\pm$1.5 \\
13  & 20080428T21:35:06 & 54584.899 & N & 6.00   &  9.0  &  28.4  &  20.9 $\pm$2.3  &  29.2  &  7.4 $\pm$0.5  &  1544 $\pm$65  &  14.5 $\pm$1.4  &  --  &    13.3 $\pm$1.4 \\
14  & 20080428T23:30:10 & 54584.980 & N & 7.20  &  11.0  &  20.2  &  18.5 $\pm$2.1  &  22.0 & 5.3 $\pm$0.4 & 982 $\pm$38 & 14.0 $\pm$1.4 & 100.0 $\pm$4.3 & 12.8 $\pm$1.4 \\
15  & 20080429T21:07:54 & 54585.881 &  N & 5.70  &  13.3  &  26.2  &  21.2 $\pm$2.4  &  26.0  &  5.2 $\pm$0.4  &  1104 $\pm$39  &  13.8 $\pm$1.4  &  --  &   12.7 $\pm$1.4 \\
16  & 20080429T22:46:16 & 54585.950 &  N & 6.70  &  7.8  &  20.1  &  18.4 $\pm$1.5  &  23.2  & 5.8 $\pm$0.4 & 1069 $\pm$13 & 13.9 $\pm$1.4 & 77.6 $\pm$1.8 & 12.8 $\pm$1.4 \\
17  &  20080501T23:45:27  &  54587.999 &  N & 6.50   &  9.0  &  26.9  &  23.2 $\pm$2.6  &  25.5  &  5.7 $\pm$0.4  &  1324 $\pm$56  &  11.8 $\pm$1.2  &  -- & 10.8 $\pm$1.2 \\
18  &  20080507T19:12:39  &  54593.800 &  N & 6.75   &  11.5  &  27.7  &  24.6 $\pm$2.9  &  40.5  &  5.2 $\pm$0.4  &  1277 $\pm$51  &  12.5 $\pm$1.3  & -- & 11.5 $\pm$1.3 \\
19  &  20080510T02:45:24 &   54596.115 & N & 9.00   &  12.6  &  29.7  &  26.1 $\pm$4.0  &  37.8  &  4.5 $\pm$0.4  &  1173 $\pm$74  &  13.5 $\pm$1.3  &  -- & 12.4 $\pm$1.3 \\
20  &  20080517T16:37:04 &   54603.692 &  N & 5.25   &  10.7  &  17.5  &  17.4 $\pm$1.6  &  26.2  &  5.7 $\pm$0.4  &  990 $\pm$21  &  12.0 $\pm$1.2  & -- & 11.0 $\pm$1.2 \\
21  &  20080519T10:36:34 &   54605.442 &  N & 6.25   &  5.9  &  37.2  &  29.2 $\pm$6.0  &  43.5  &  4.4 $\pm$0.4  &  1286 $\pm$149  &  12.3 $\pm$1.2  &  -- & 11.3 $\pm$1.2 \\
22  & 20080519T12:25:58 &  54605.518 &  N & 7.50  &  9.6  &  28.0  &  22.6 $\pm$3.0  &  29.0 & 4.4 $\pm$0.4 & 993 $\pm$42 & 12.2 $\pm$1.2 & 80.4 $\pm$4.6 & 11.2 $\pm$1.2 \\
23  &  20080521T02:32:06 &   54607.106  &  N & 6.50   &  8.4  &  22.9  &  18.9 $\pm$1.8  &  24.5  &  7.4 $\pm$0.5  &  1398 $\pm$39  &  12.3 $\pm$1.2  & -- & 11.3 $\pm$1.2 \\
24  &  20080522T13:55:32 &   54608.580  &  N & 6.50   &  8.3  &  25.8  &  21.5 $\pm$2.1  &  26.0  &  6.1 $\pm$0.4  &  1310 $\pm$43  &  12.1 $\pm$1.2  & -- & 11.1 $\pm$1.2 \\
25  &  20080607T07:02:53  &  54624.294   &  N & 7.00   &  17.5  &  36.2  &  31.5 $\pm$6.1  &  40.0  &  4.3 $\pm$0.4  &  1355 $\pm$135  & 13.6 $\pm$1.4 & -- & 12.5 $\pm$1.4 \\
26  &  20080609T05:50:35  &  54626.243   &  N & 7.00   &  10.5  &  37.2  &  35.3 $\pm$6.7  &  47.0  &  4.7 $\pm$0.4  &  1660 $\pm$174  & 14.2 $\pm$1.4 & -- & 13.0 $\pm$1.4 \\
27  &  20080611T00:38:29 &   54628.027  &  N & 5.50   &  8.7  &  23.9  &  21.5 $\pm$1.9  &  24.5  &  6.4 $\pm$0.4  &  1374 $\pm$36  &  15.5 $\pm$1.5  & -- & 14.2 $\pm$1.5 \\
28  &  20080612T05:06:52 &   54629.213  &  N & 8.00   &  10.0  &  31.8  &  27.5 $\pm$5.0  &  36.0  &  4.3 $\pm$0.4  &  1184 $\pm$107  &  15.9 $\pm$1.6 & -- & 14.6 $\pm$1.6 \\
29  &  20080613T12:33:43 &   54630.523  &  N & 7.50   &  7.1  &  21.5  &  20.7 $\pm$2.0  &  26.5  &  6.0 $\pm$0.4  &  1241 $\pm$36  &  16.0 $\pm$1.6  & -- & 14.7 $\pm$1.6 \\
30  &  20080615T00:12:29 &   54632.009  &  N & 7.00   &  7.5  &  30.2  &  27.3 $\pm$4.3  &  38.0  &  4.1 $\pm$0.3  &  1118 $\pm$94  &  16.4 $\pm$1.6  & -- & 15.0 $\pm$1.6 \\
31  &  20080616T20:03:02 &   54633.836  &  N & 9.00   &  10.0  &  29.5  &  25.5 $\pm$3.4  &  39.0  &  4.0 $\pm$0.3  &  1021 $\pm$60  &  16.4 $\pm$1.6 & -- & 15.0 $\pm$1.6 \\
32  &  20080617T21:48:55 &   54634.909  &  N & 8.00   &  10.1  &  30.9  &  27.1 $\pm$3.6  &  38.5  &  3.9 $\pm$0.3  &  1055 $\pm$61  &  18.4 $\pm$1.8 & -- & 16.9 $\pm$1.8 \\
33  &  20080717T08:25:41 &   54664.351  & \textbf{Y} & 2.25   &  3.4  &  10.3  &  8.9 $\pm$0.6  &  8.5  &  9.6 $\pm$0.6  & 853 $\pm$8 & 4.6 $\pm$0.5 & -- & 4.2 $\pm$0.5 \\
34  &  20080717T11:41:00 &   54664.487  &  N & 3.75   &  2.4  &  6.7  &  7.7 $\pm$0.6  &  8.5 & 6.3 $\pm$0.4 & 486 $\pm$7 & 4.6 $\pm$0.4 & 110.7 $\pm$3.6 & 4.2 $\pm$0.4 \\
35  &  20080719T08:38:07 &   54666.360  &  \textbf{Y} & 3.00  &  3.7  &  10.9  &  9.6 $\pm$0.7  &  9.5  &  9.5 $\pm$0.6  & 911 $\pm$9 & 3.9 $\pm$0.4 & -- & 3.6 $\pm$0.4 \\
36  &  20080729T18:49:21 &   54676.784  &  N & 4.00   &  7.2  &  17.1  &  17.7 $\pm$1.6  &  17.0  &  5.1 $\pm$0.4  &  902 $\pm$11  &  3.9 $\pm$0.4  &  -- & 3.6 $\pm$0.4 \\
37  &  20080814T10:13:10 &   54692.426  &  N & 5.00   &  6.2  &  17.5  &  16.7 $\pm$1.5  &  16.5  &  4.9 $\pm$0.4  &  817 $\pm$5  &  3.6 $\pm$0.4  &  -- & 3.3 $\pm$0.4 \\
38  &  20080903T06:50:42  &  54712.254   &  N & 3.25   &  6.6  &  18.8  &  13.1 $\pm$1.0  &  16.0  &  7.4 $\pm$0.5  &  972 $\pm$6  &  3.5 $\pm$0.3 & -- & 3.2 $\pm$0.3 \\
39  &  20080904T08:37:38  &  54713.359   &  N & 4.00   &  6.8  &  19.2  &  13.6 $\pm$0.9  &  17.2  &  6.6 $\pm$0.4  &  895 $\pm$5  &  3.3 $\pm$0.3  & -- & 3.0 $\pm$0.3 \\
40  &  20080908T09:27:36  &  54717.394   &  N & 4.25   &  5.7  &  15.6  &  11.2 $\pm$0.7  &  15.2  &  8.4 $\pm$0.5  &  940 $\pm$5  &  3.1 $\pm$0.3  & -- & 2.8 $\pm$0.3 \\
41  &  20080914T00:28:23 &   54723.020  &  N & 4.75   &  4.9  &  11.9  &  9.8 $\pm$0.7  &  12.2  &  7.8 $\pm$0.5  &  768 $\pm$5  &  1.9 $\pm$0.2  &  -- & 1.7 $\pm$0.2 \\
42  &  20080917T16:54:30 &   54726.704  &  \textbf{Y} & 5.5   &  4.8  &  16.8  &  11.2 $\pm$0.8 & 13.5 & 8.0 $\pm$0.5 & 898 $\pm$7 & 1.2 $\pm$0.1 & -- & 1.1 $\pm$0.1 \\
INTEGRAL \\ 
   (JEM-X: 3-25 keV) &  &  &  &  $\pm$1  &  &   &  &  $\pm$5  &    &    &    &  &   \\
1 & 20080406T12:11:12 & 54562.50778  &  &  3  &  & 9.7  $\pm$1.2 & 9.3 $\pm$1.9 &   20 & 9.2 $\pm$1.8  &    851 $\pm$170 & 5.0 $\pm$0.6  &  --  &  4.6 $\pm$0.6 \\
2 & 20080406T18:40:43 & 54562.77827  &  &  4  &  & 13.4 $\pm$3.2 & 13.4 $\pm$2.7 &  60 & 5.9 $\pm$1.2  &    788 $\pm$160 & 5.1 $\pm$0.6  &  151 $\pm$30  &    4.7 $\pm$0.6 \\
3* & 20080408T13:36:28 & 54564.56699  &  &  7  &  & 10.7 $\pm$3.0 & 10.4 $\pm$5.0 &  25 & 3.8 $\pm$1.9  &    396 $\pm$150 & 5.9 $\pm$0.7   &  --  &   5.4 $\pm$0.7 \\
4 & 20080411T16:29:09 & 54567.68691  &  &  5  &  & 22.8 $\pm$3.5 & 21.9 $\pm$4.4 &  75 & 4.2 $\pm$0.8   &    921 $\pm$190 & 8.9 $\pm$1.0   &  --  &   8.2 $\pm$1.0 \\
5 & 20080411T19:45:40 & 54567.82338  &  &  5  &  & 25.5 $\pm$6.1 & 25.0 $\pm$5.1 & 100 & 5.1 $\pm$1.0  &   1274 $\pm$260 & 9.0 $\pm$1.0   &  83 $\pm$20  &   8.3 $\pm$1.0 \\
6 & 20080414T10:13:31 & 54570.42605  &  &  4  &  & 8.0? $\pm$3.3+ & $>7.7$ &  $>25$ & 4.4 $\pm$0.8 &  $>337$ & 11.1 $\pm$1.2  &  --  &   10.2 $\pm$1.2 \\
7 & 20080414T13:04:58 & 54570.54512  &  &  6  &  & 17.9 $\pm$3.5 & 17.8 $\pm$3.6 &  90 & 5.0 $\pm$1.0  &  889 $\pm$180 & 11.2 $\pm$1.2  &  130 $\pm$30  &    10.3 $\pm$1.2 \\
8 & 20080417T10:52:07 & 54573.45286  &  &  3  &  & 13.8 $\pm$3.2 & $>13.0$ &  40+ & 4.2 $\pm$0.8 &  $>548$  & 12.3 $\pm$1.3  &  --  &    11.3 $\pm$1.3 \\
9 & 20080417T13:21:24 & 54573.55653  &  &  7  &  & 23.9 $\pm$5.0 & 23.0 $\pm$4.6 &  80 & 4.2 $\pm$0.8 & 968 $\pm$200 & 12.3 $\pm$1.3  &   114 $\pm$25  &   11.3 $\pm$1.3 \\
10 & 20080420T10:40:58 & 54576.44512  &  &  9  &  & 24.5 $\pm$5.9 & 24.1 $\pm$4.8 &  100 & 5.1 $\pm$1.0 & 1228 $\pm$260 & 11.3 $\pm$1.2  & (76 $\pm$20)  &   10.4 $\pm$1.2 \\
11 & 20080420T12:46:37 & 54576.53237  &  &  4.5  &  &  33.1 $\pm$11.3 & 29.7 $\pm$5.9 & 75 & 4.0 $\pm$0.8 & 1186 $\pm$240 & 11.3 $\pm$1.2  & 72 $\pm$15  &   10.4 $\pm$1.2 \\
12 & 20080818T04:59:14 & 54696.20780  &  &  5  &  & 9.3  $\pm$1.4 & 9.3 $\pm$1.9 &  50 & 7.9 $\pm$1.6  &    731 $\pm$150 & 3.3 $\pm$0.4  &  --  &    3.0 $\pm$0.4 \\
13 & 20080830T03:39:54 & 54708.15201  &  &  2  &  & 7.4  $\pm$1.0 & $>7.3$ &  $>30$ & 9.2 $\pm$1.8 &   $>667$ & 3.6 $\pm$0.5  &  --  &    3.3 $\pm$0.5 \\
14 & 20080902T03:33:43 & 54711.14841  &  &  1  &  & 14.1 $\pm$1.7 & 13.7 $\pm$2.8 &  50 & 5.5 $\pm$1.1  &    753 $\pm$150  &  3.5 $\pm$0.5  &  --  &    3.2 $\pm$0.5 \\
\hline
\end{tabular}
\\
\small  All fluxes $F_{\rm peak}$ and $F_{\rm pers}$, as well as fluences are bolometric (0.1--200 keV), but for the SuperAGILE burst $F_{\rm peak}$ and Fluence are between 17--25 keV. \\ 
\small $^{a}$ Rise time. 
\small $^{b}$ Exponential decay time. 
\small $^{c}$ $\tau \equiv \rm Fluence/F_{\rm peak}$. 
\small $^{d}$ $\alpha$: see text section 2.5. 
\small $^{e}$ $\gamma \equiv F_{\rm pers}/F_{\rm peak,Max}$;  
$F_{\rm peak,Max}$ is the highest burst peak flux, here $F_{\rm peak}=10.9\times10^{-8}$~\ergcs 
of the \S\ burst. 
\small $^{*}$ indicates the same burst simultaneously observed by JEM-X (\#3) and RXTE (\#1).
\end{minipage}
}
}
\vfil}
\end{table*}

\end{document}